\documentclass{aa}  

\usepackage{graphicx}
\usepackage{tabularx}
\usepackage{booktabs}
\usepackage[normalem]{ulem} 
\usepackage{amsmath}    
\usepackage{amssymb}    
\usepackage{threeparttable}
\usepackage{color}

\usepackage{txfonts}
\begin{document} 

\title{Observations of meteors in the Earth's atmosphere:\\ 
Reducing data from dedicated double-station wide-angle cameras}
\titlerunning{Observations of meteors in the Earth's atmosphere}

\author{A. Margonis\inst{1}, A. Christou\inst{2} and J. Oberst\inst{1,3}}


\institute{Department of Geodesy and Geoinformation Science, Technische Universit\"{a}t Berlin, Strasse des 17.~Juni 135, Berlin, Germany\\
\email{anastasios.margonis@tu-berlin.de}
\and
Armagh Observatory, College Hill, Armagh BT61 9DG, UK\\
\email{aac@arm.ac.uk}
\and
Germany Aerospace Center, Institute of Planetary Research, Rutherfordstr. 2, 12489 Berlin, Germany\\
\email{juergen.oberst@dlr.de}
          }

\date{Received 2018; accepted 2018}

 
  \abstract{Meteoroids entering the Earth's atmosphere can be observed as meteors, thereby providing useful information on their formation and hence on their parent bodies. We developed a data reduction software package for double station meteor data from the SPOSH camera, which includes event detection, image geometric and radiometric calibration, radiant and speed estimates, trajectory and orbit determination, and meteor light curve recovery. The software package is designed to fully utilise the high photometric quality of SPOSH images. This will facilitate the detection of meteor streams and studies of their trajectories. We have run simulations to assess the performance of the software by estimating the radiants, speeds, and magnitudes of synthetic meteors and comparing them with the a priori values. The estimated uncertainties in radiant location had a zero mean with a median deviation between 0.03$^{\circ}$ and 0.11$^{\circ}$ for the right ascension and 0.02$^{\circ}$ and 0.07$^{\circ}$ for the declination. The estimated uncertainties for the speeds had a median deviation between 0.40 and 0.45 km s$^{-1}$. The brightness of synthetic meteors was estimated to within +0.01m. We have applied the software package to 177 real meteors acquired by the SPOSH camera. The median propagated uncertainties in geocentric right ascension and declination were found to be of 0.64$^{\circ}$ and 0.29$^{\circ}$, while the median propagated error in geocentric speed was 1.21 km $s^{-1}$.}


   \maketitle

\section{Introduction}
\label{intro}

Observations of meteors in the Earth's atmosphere shed light on the properties of the population of meteoroids intercepting the orbit of our planet. The study of the temporal and spatial distribution of meteors requires sensitive optical systems that are able to monitor the night sky. Double station observations (i.e. observations of two cameras from different positions) are required to determine the trajectories and orbit parameters of the meteors. 

While algorithms for meteor data reduction are well established in the literature \citep{Ceplecha1987,Trigo2004,Weryk2008,Jenniskens2011}, every camera may require an analysis system to account for the specific capabilities of the camera.

For observations of meteors in recent years, our team has used the Smart Panoramic Optical Sensor Head (SPOSH) camera \citep{Oberst2011}. The instrument features a highly sensitive CCD chip that delivers images of high photometric quality. With the wide-angle lens, the camera easily captures several hundreds of stars in one image, which requires sophisticated geometric calibration procedures.

To process the data from SPOSH, we developed a comprehensive software package, which allows us to carry out camera calibration, meteor detection, meteor trajectory determination, meteor photometric modelling, and orbit reconstruction. In this paper, we describe this software and demonstrate and assess its performance on synthetic and actual meteor data. 

\section{The SPOSH camera}
\label{sposh_camera}

The SPOSH camera was designed to image faint transient noctilucent phenomena, such as aurorae, electric discharges, meteors, or impact flashes on dark planetary hemispheres from an orbiting platform \citep{Oberst2011}. The camera is equipped with a highly sensitive back-illuminated 1024$\times$1024 CCD chip and has a custom-made optical system of high light-gathering power with a wide field of view (FOV) of 120$\times$120$^{\circ}$. The SPOSH camera system is accompanied by a sophisticated digital processing unit (DPU) designed for real-time image processing and communication with a spacecraft. Owing to the all-sky coverage and excellent radiometric and geometric properties  of the camera, a large number of meteors can be obtained for reliable event statistics.  

For outdoor tests and meteor monitoring, the camera is typically mounted on a tripod pointed vertically up at the sky taking one image every 2 s. For the determination of the meteor velocity, a mechanically rotating shutter with a known frequency is mounted in front of the camera lens. The shutter consists of two blades and has a rotating frequency of 250 RPM resulting in an exposure time of 0.06 s for every shutter opening. Double-station observations have been carried out routinely providing a large dataset of meteor images.

\section{Data reduction}
\label{data_reduction}
\begin{figure}\centering
  \includegraphics[width=0.47\textwidth]{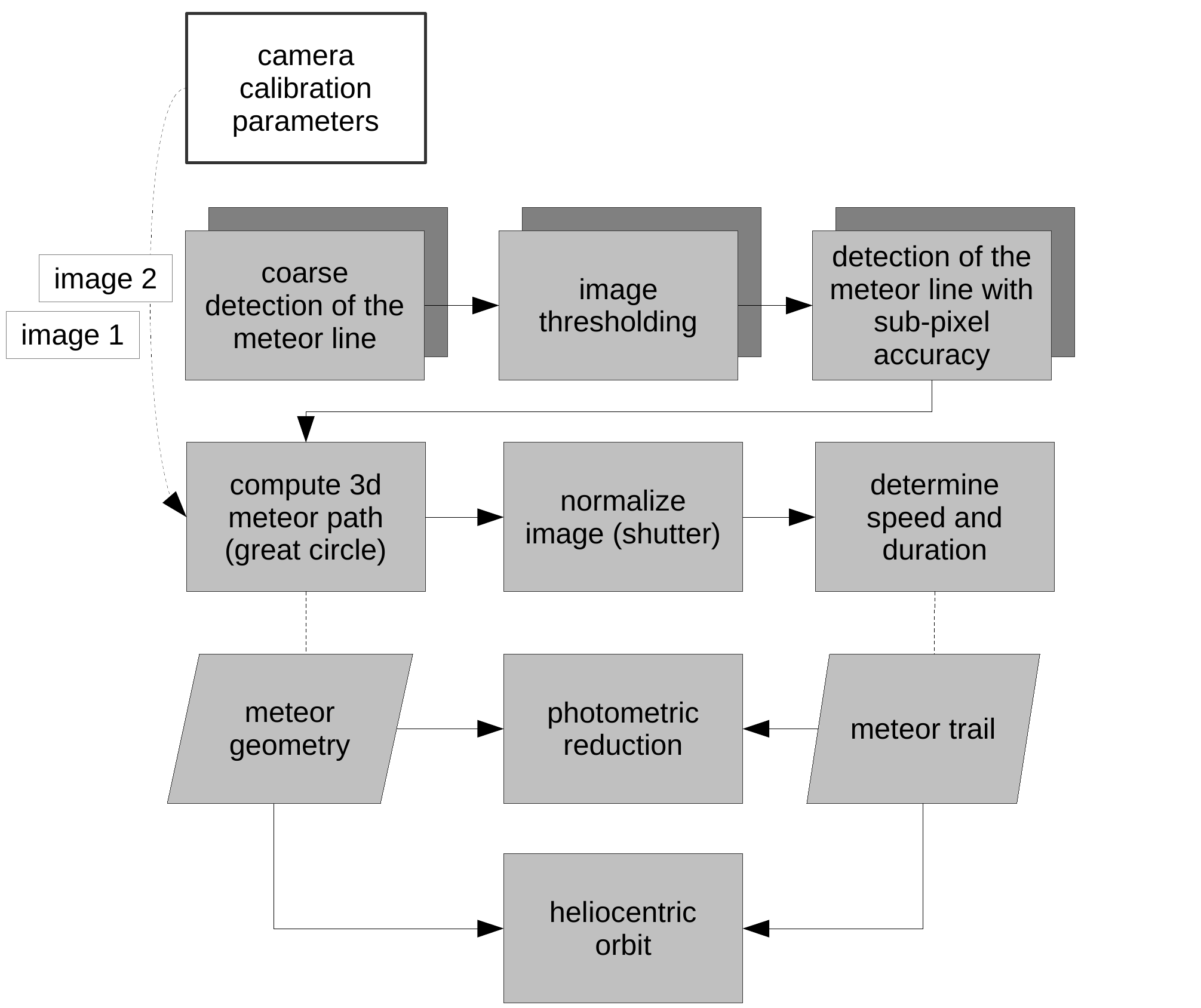}
  \caption{Flow chart showing the different modules of the software package. The camera calibration software is used as a stand-alone program and in the flowchart is depicted as a rectangle with a white background.}
  \label{fig:flowchart}
\end{figure}

The reduction of the meteor data is performed by different software modules. The calibration software SPOSHCalib is a stand-alone software for the geometric calibration of SPOSH images \citep{Elgner2006}. The trajectory determination module is based on the MOTS software \citep{Koschny2002}, which was initially modified to process SPOSH data \citep{Maue2006}. All modules were developed anew within the scope of this study. The interaction between the different modules can be seen in Figure \ref{fig:flowchart}.

\subsection{Meteor detection}
\label{meteor_detection}

Unlike video cameras, where a meteor only spends a fraction of their trajectory in each frame, exposures longer than one second often capture the whole meteor (e.g. a Perseid) in one image. The meteor detection algorithm that we used is based on the Hough transform technique for extracting linear features within images. This method has been used to detect meteors in photographic image data by previous authors \citep{Trayner1996,Gural1997}. 

The algorithm that we developed first generates 8 bit difference images between three consecutive frames, removing the background and highlighting only short temporal variations. Possible non-relevant information depicted in the margins of the images (e.g. surrounding mountains and man-made structures), typical within large FOVs, are removed by applying a circular mask. Background noise and stellar scintillation are filtered out by first applying an empirical threshold and then a median filter to the image, thus reducing the overall computation time of the algorithm. Each line, represented by a combination of $\rho$ and $\theta$, passing through each of the remaining pixels contributes to the parameter space $H(\rho, \theta)$, known also as voting space, by adding the value $A_{xy}$=1, 
\begin{equation}
 H(\theta,\rho) = \sum_{x}\sum_{y}A_{xy}\delta(\rho, [\rho'])
,\end{equation}
where $\rho$ is the distance from the origin to the closest point on the straight line, and $\theta$ is the angle between the $x$ axis and the line connecting the origin with that closest point. The square brackets [ ] indicate rounding to the nearest integer and the normal representation of a line is
\begin{equation}
 \rho'=xcos\theta+ysin\theta,
\end{equation} 
with
\begin{align}
\delta(\rho, [\rho'])= 
 \begin{cases}
   1, \quad \rho = \rho' \nonumber\\
   0, \quad otherwise.      
 \end{cases}
\end{align} 
The event detection algorithm is triggered each time a certain threshold value is exceeded. This value is compared against the maximum value found in the parameter space of each image and represents the number of pixels lying on a line in the image space. 

\begin{figure}\centering
  \includegraphics[width=0.35\textwidth]{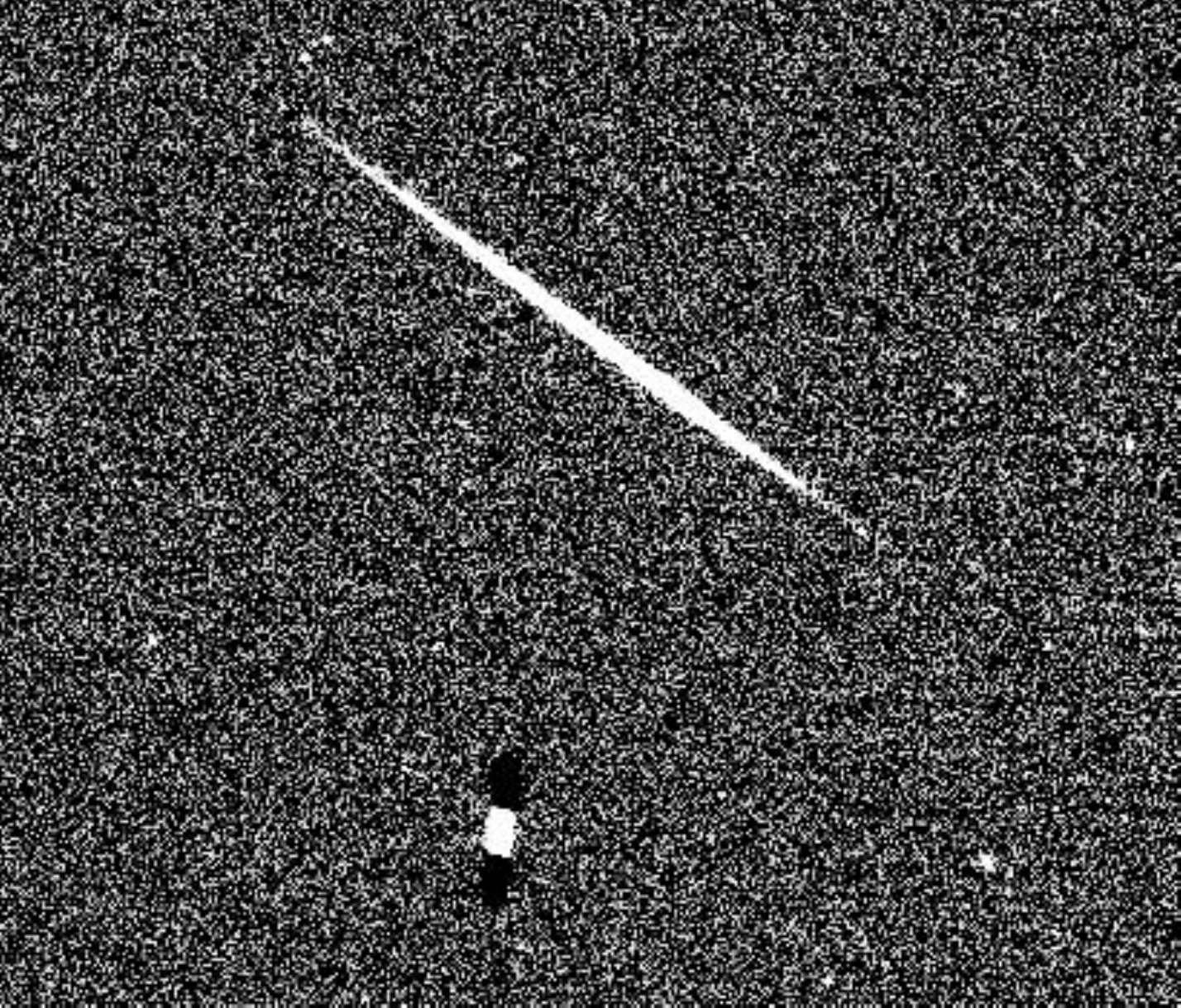}
  \caption{Difference image showing a meteor trail and an airplane with its characteristic negative-positive-negative pattern in the lower part of the image.}
  \label{fig:mask_rot}
\end{figure}

Several criteria are used to mitigate the effect of false detections. Slow moving objects, such as airplanes and satellites, appear with a characteristic negative-positive-negative pattern in the difference images (Fig.~\ref{fig:mask_rot}). This pattern is compared against a predefined signal by the user, simulating the path of an airplane projected on the image plane. In this way, events appearing in three consecutive images moving with low apparent angular velocities of 0.6$^{\circ}$ $>$ v$_{ang}$ $>$ 2.2$^{\circ}$ are rejected as slow-moving objects. This condition also affects meteors appearing close to their radiant position and/or close to the horizon. For every event, its time of occurrence, the central position of the line, and its direction within the image are saved together with the object name (meteor, slow-moving object, or star) in a text file. 
A quality parameter $q_{md}$ was introduced to determine the threshold value for the Hough transform. The value of the threshold should ideally detect all meteors in the image data when applied to a meteor detection algorithm. At the same time slow-moving objects and random noise patterns resembling lines should be filtered out. To select a suitable value for this parameter, we balance the number of false positives against the number of meteors the algorithm failed to detect (false negatives) and the processing time it takes for the algorithm to scan the images. The value is computed, after applying various weights to the observed quantities, as follows:
\begin{equation}\label{eq:quality}
 q_{md} = \frac{p_1w_1-p_2w_2-p_3w_3}{100w_1-8w_3}
,\end{equation}
where $p_{1}$ is the percentage of the detected meteors, $p_{2}$ the percentage of false detections, $p_{3}$ the processing time in minutes, and $w_{1}$, $w_{2}$, and $w_{3}$ the respective weights. The quality parameter is scaled to values between 0 and 1 (Eq. \ref{eq:quality}). The maximum value ($q_{md}$=1) for a threshold is reached when all meteors are detected ($p_{1}$=100), with no false detections ($p_{2}$=0), within a user-defined processing time. 

\begin{figure}\centering
  \includegraphics[width=0.5\textwidth]{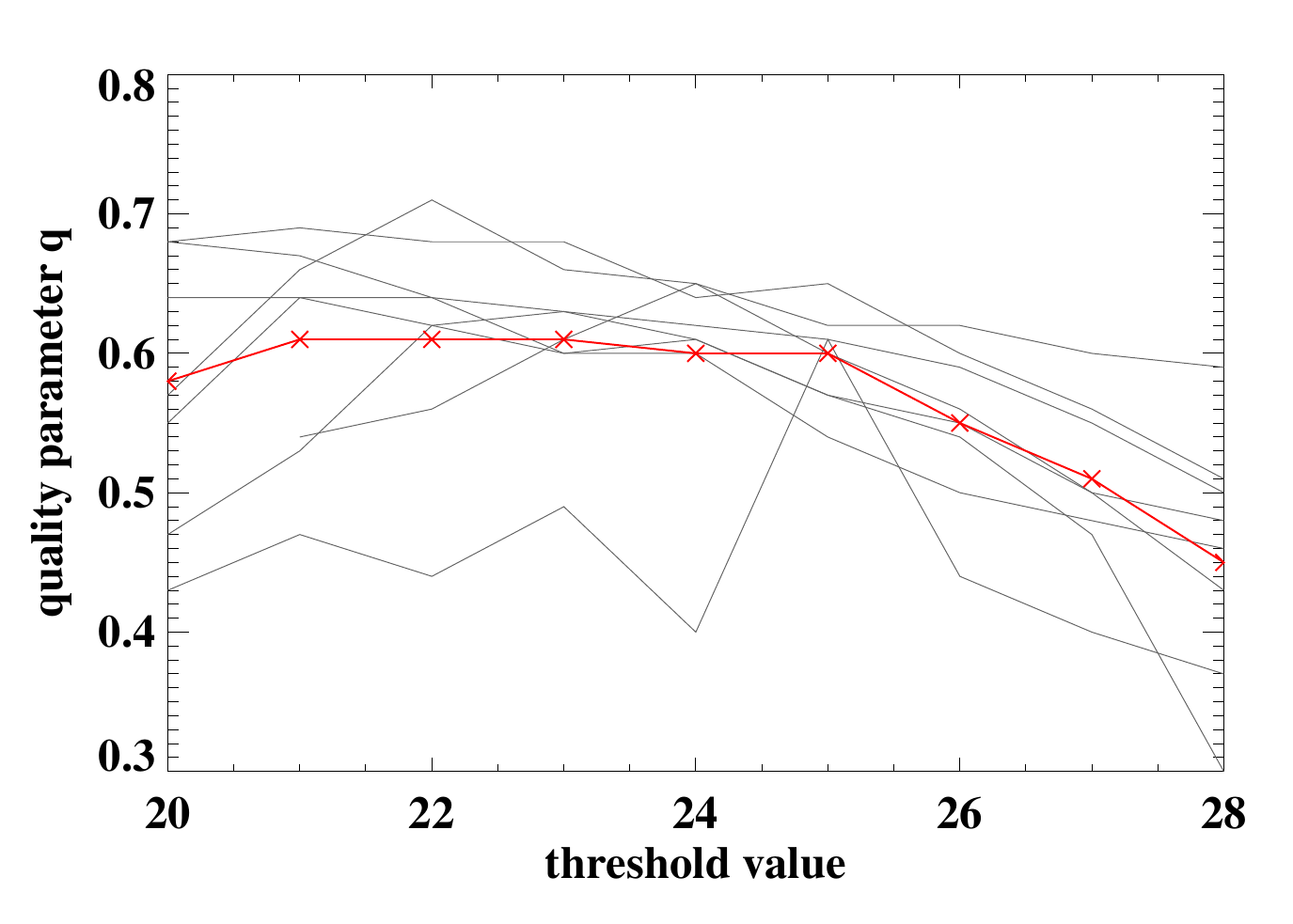}
  \caption{Quality parameter values computed from 8 h of image data with respect to different threshold values. The faint lines show the values of the quality parameter $q_m$ for each of the 8 datasets while the red line shows the mean value of the quality parameter for the different threshold values.}
  \label{fig:thres_plot}
\end{figure}

We tested the performance of our algorithm using various threshold values and applying these to $>$ 14,000 images corresponding to eight hours of data from two observing sites. The results were compared with meteors identified after visual inspection of the images. The highest value of the quality parameter for this dataset was found for a threshold of 23 with $w_{1}$:0.6, $w_{2}$:0.3, and $w_{3}$:0.1. The threshold value corresponds to the highest number of pixels lying on a line in a given image. Applying these parameters, 70\% of the visually identified meteors were successfully detected by the algorithm (true positives), while 15\% of the detected events were false detections (false positives). Figure~\ref{fig:thres_plot} shows the calculated quality parameter for our dataset. High values are computed from data with a relative high signal-to-noise ratio in terms of detected meteors and false detections. This performance of the algorithm can be achieved under favourable weather conditions.

\subsection{Astrometry}

\subsubsection{Camera calibration}
\label{calibration}

The geometric calibration of the camera is performed by the SposhCalib software in a semi-automatic process using standard stars in the SPOSH images \citep{Elgner2006}. Stars are ideal calibration targets owing to their high abundance in the images and the precise knowledge of their position at a given time. The SPOSH images may feature up to several thousand stars, which are on average equally distributed over the whole image except image corners. By comparing the actual stars in the image with their expected positions based on a priori information about pointing and interior camera parameters, these parameters can be updated in a least-squares fashion. This provides an accurate knowledge of the interior, i.e. focal length and geometric distortion, and the exterior orientation (pointing) parameters. The coordinates of the stars are taken from the Tycho-2 and Hipparcos star catalogues \citep{ESA1997}.

The transformation equations between the image coordinate system (\textit{x,y}) and the camera coordinate system (\textit{X$_{cam}$,Y$_{cam}$,Z$_{cam}$}) are described applying an equidistant camera model \citep{Ray1994},
\begin{align} \label{eq:equid}
 & X_{cam} = \frac{x_c}{\sqrt{x_c^2+y_c^2}}sin(\sqrt{x_c^2+y_c^2}), \nonumber \\
 & Y_{cam} = \frac{y_c}{\sqrt{x_c^2+y_c^2}}sin(\sqrt{x_c^2+y_c^2}), \\
 & Z_{cam} = \cos\left(\sqrt{x_c^2+y_c^2} \right) \nonumber. 
\end{align}
A high number of standard stars is achieved by performing initially a pre-calibration with the help of at least six reference stars selected by the user. The pre-calibration step provides approximate values for the unknown parameters. After this step, a global calibration is performed using all point sources identified as standard stars in the image. 

The SPOSH images show significant radial and non-symmetrical distortion, mathematically expressed as
\begin{equation} \label{eq:radial}
 \Delta r_{rad} = A_1r^2 + A_2r^4 + A_3r^6
\end{equation}
\begin{align} \label{eq:tang}
 & \Delta x_{tan} = B_1(r^2+2x^2)+2B_2xy, \nonumber \\
 & \Delta y_{tan} = B_2(r^2+2x^2)+2B_1xy
\end{align}
\begin{align} \label{eq:affine}
 & \Delta x_{aff} = C_1x, \nonumber \\
 & \Delta y_{aff} = -C_1y
\end{align}
\begin{align} \label{eq:shear}
 & \Delta x_{sh} = C_2y, \nonumber \\
 & \Delta y_{sh} = C_2x
\end{align}
with 
\begin{equation}
 r = \sqrt{x^2+y^2} \nonumber .
\end{equation}
The equations above describe the radial (Eq. \ref{eq:radial}) and non-symmetrical distortions (Eq. \ref{eq:tang}) and the deviations of the image coordinate system from an orthogonal, uniformly scaled coordinate system (Eqs. \ref{eq:affine}, \ref{eq:shear}). The outer and inner orientation of the camera and the distortion parameters introduced by the lens are determined by fitting a 6$^{th}$-order polynomial function. These distortion terms are added directly to the pixel coordinates of the stars. 

The average residual error for the star positions after the calibration is usually less than 0.25 pixel or 1.68$\arcmin$ and usually consistent over the whole image. The displacement $\Delta$$x_{ij}$ and $\Delta$$y_{ij}$ in image coordinates due to radial distortion are stored in two separate TIFF files. These files serve as look-up tables in the subsequent steps providing the undistorted position of each pixel.

\subsubsection{Meteor path on the image plane}
\label{meteor_line}

The projection of a trajectory of a meteor on an image plane can be seen as a meteor trail. By extending the trajectory before and after the luminous path, a line can be defined on the image plane representing the projection of that extended path. Once the line is defined in both images, its radiant can be determined (Section \ref{meteor_geometry}).

\begin{figure}\centering
  \includegraphics[width=0.47\textwidth]{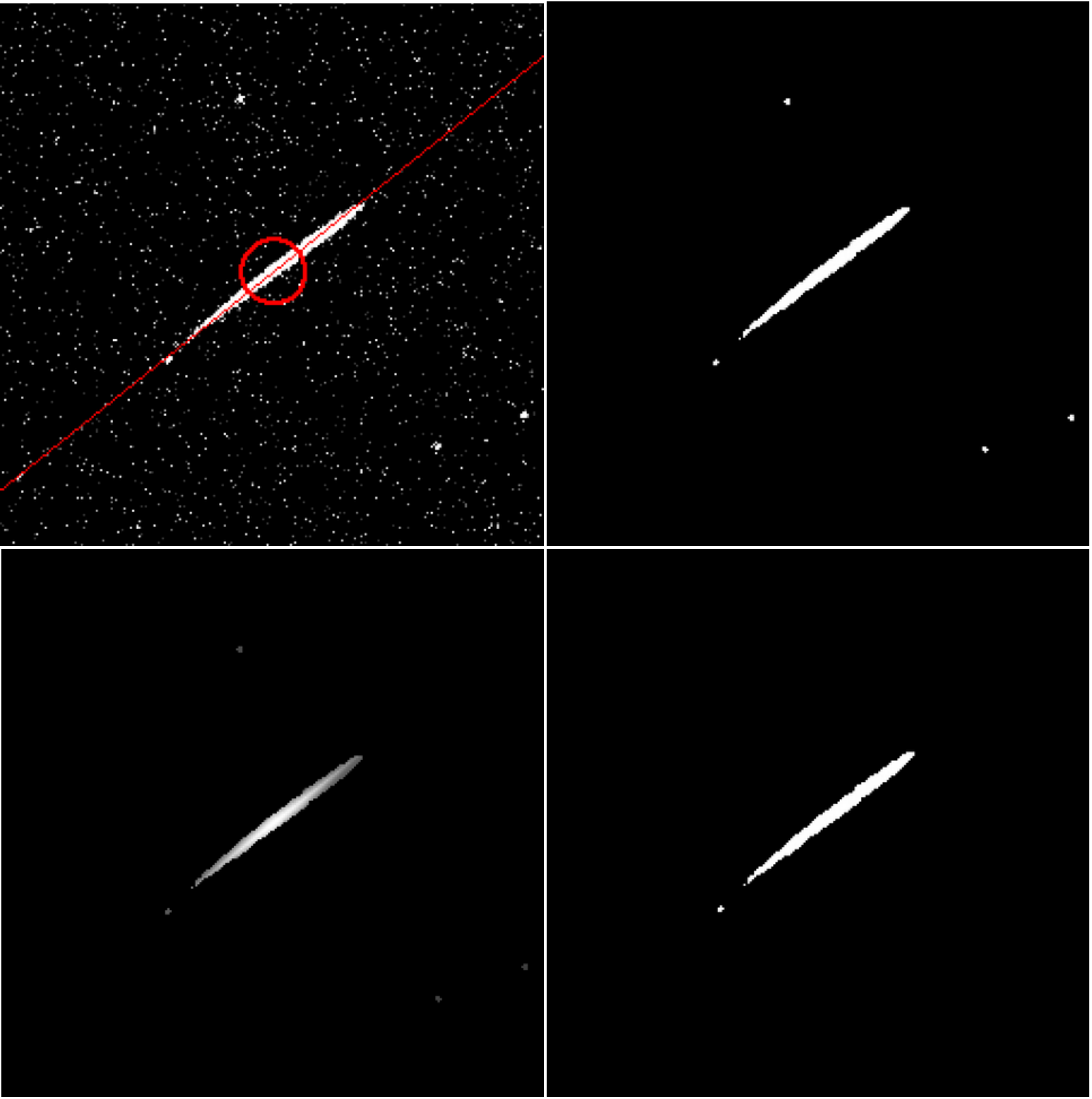}
  \caption{Plots showing the consecutive processing steps of a meteor image for the removal of structures not belonging to the meteor trail. From top left to lower right: detected meteor line and pixel with maximum votes (red) in thresholded image, median filter, computed coefficients for each pixel. High intensities represent meteor pixel and selected meteor pixel.}
  \label{fig:pixonline}
\end{figure}

In order to speed up the process of defining the meteor line, a threshold is applied to each raw meteor image. The threshold is defined at 2$\sigma$ of the noise level. The use of a relative low threshold ensures that fainter pixels belonging to the meteor trail are considered in the computation of the line. The line parameters defined in the meteor detection procedure (Section \ref{meteor_detection}) are used to remove unwanted features in each meteor image, considering the proximity of each pixel to the detected line, its distance to the pixel with the maximum votes in the Hough transform, and its intensity value (Fig.~\ref{fig:pixonline}). 

The positions of the remaining pixels are corrected for radial distortion by retrieving pixel-offset values from the look-up tables generated in the calibration step (Section \ref{calibration}). Owing to the equidistant projection model used by the lens system of the camera, perspective distortions in the image are evident that deflect the path of objects moving along a great circle from a straight line to a curved line. To efficiently detect linear features in the image, pixel coordinates are converted from an equidistant to a gnomonic projection, where straight lines in space preserve their straightness when projected on the image plane. 

\begin{figure}\centering
  \includegraphics[width=0.4\textwidth]{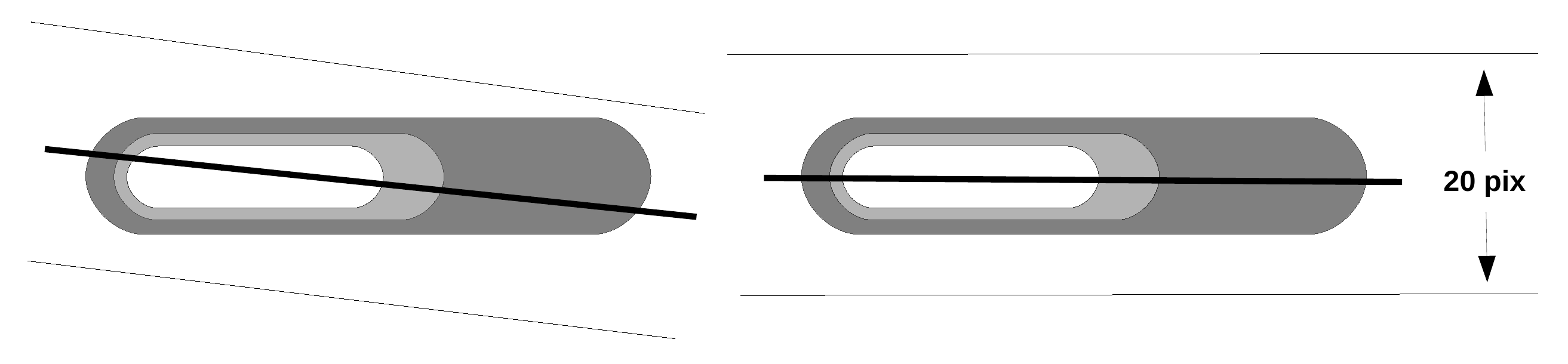}
  \caption{Simplified meteor example represented by three intensity levels with the dark grey area corresponding to low dn values. The line intersecting the meteor in the left example has the highest value in voting space, while the right line produces the highest ratio and it is the desired outcome. A buffer zone with a width of 20 pixel parallel to the determined line is depicted by the two parallel thin lines.}
  \label{fig:diag_met_line}
\end{figure}

The line along which the meteor is moving is computed by applying a customised Hough transform. As an input, we use the corrected image coordinates (in sub-pixel accuracy) belonging to the meteor trail. Lines running diagonal to the meteor trail results into a higher value in voting space than those parallel to the meteor motion, since more pixels lie along the diagonal line  (Fig.~\ref{fig:diag_met_line}). We handle this effect as follows: First we apply a Hough transform to determine the top 20 lines intersecting the highest number of pixels. Then we perform a weighted Hough transform considering the intensity values. Unlike the standard Hough transform method, which searches for the line with the maximum votes $V$ in parameter space, we defined a ratio coefficient calculated as the sum of intensity values $I$ with respect to the number of pixels that are $\sqrt{2}/2$ pixels apart from each line parameter combination. A distance of $\sqrt{2}/2$ pixel is needed to identify which pixels lie on the line since the line does not cross the pixel centre (defined at 0.5 pix),

\begin{equation}
 V_{max}(\theta,\rho) = \sum\limits_{i=1}^n I_{i}/n
.\end{equation}

The best-fitting line is defined as the line with the highest ratio. Since the point spread function (PSF) of an imaging system spreads the light of point sources to neighbouring pixels, the light emitted by a meteor also spreads to pixels located perpendicular to its motion. In order to account for the signal within these pixels, we define a buffer zone of 10 pixels perpendicular to the best line computed.

Occasionally, residual features may be located along the buffer zone. As a result, these remaining pixels affect the determination of the meteor line. In order to remove these features, the consecutive pixel-to-pixel distances are determined revealing gaps between features. Distances higher than a threshold indicate different pixel entities, where entity is a feature consisting of at least two neighbouring pixels; for example, the meteor trail or meteor segment is such an entity. Assuming that the meteor entity has the maximum number of pixels, we remove all secondary features from the line zone. Finally, the meteor line is determined using weighted least squares. The line parameters (slope plus intercept) and the middle point of the meteor trail, defined as the median of the chosen pixel coordinates, are saved in a text file. 

\subsubsection{Transformation to the spatial trajectory of the meteor}
\label{image 2 space}
From the estimated parameters of the meteor line, the underlying image points are generated in sub-pixel accuracy and transformed from the gnomonic back to an equatorial projection. The pixel coordinates $x_{c}$, $y_{c}$, are normalised using the parameters of the interior orientation of the camera, i.e. (Section~\ref{calibration})
\begin{equation}\label{camera_norm}
\begin{aligned} 
 x_n &= \frac{ (x_{c}-x_p)p_x }{f}, \\
 y_n &= \frac{ (y_{c}-y_p)p_y }{f}
\end{aligned}
,\end{equation}
where $x_p$, $y_p$ are the intersection of the optical axis with the image plane (principal point), \textit{p$_x$}, \textit{p$_y$} the pixel size, and \textit{f} the focal length of the camera system. The points are first projected to the camera coordinate system ($\mid$$\vec{x}$$\mid$=1) using the equidistant projection equations. The vectors are then transformed to the local (horizontal) coordinate system, 
\begin{equation}
 \mathbf{x}_{hor} = 
 \begin{pmatrix}
  x_{hor} \\
  y_{hor} \\
  z_{hor}
 \end{pmatrix}
  = R_{\omega\phi\kappa} \cdot 
 \begin{pmatrix}
  x_{cam} \\
  y_{cam} \\
  z_{cam}
 \end{pmatrix}
,\end{equation}
where $R_{\omega\phi\kappa}$ is the 3d rotation matrix that relates the camera to the local coordinate system. Finally, the pointing vectors are transformed to the common Earth-centred, Earth-fixed (ECEF) coordinate system. The z-axis becomes parallel to the north pole by rotating the local system by an angle $90 - \phi_{geo}$ around the $x$-axis, with $\phi_{geo}$ the geocentric latitude of the camera location. The $x$-axis aligns with the direction of the prime meridian after rotating the system around the $z$-axis by an angle $\lambda_{geo}$, where $\lambda_{geo}$ the geocentric longitude of the camera location,
\begin{equation}
 \mathbf{x}_{geo} = 
 \begin{pmatrix}
  x_{geo} \\
  y_{geo} \\
  z_{geo}
 \end{pmatrix}
  = R_{\lambda_{gd} \phi_{gd}} \cdot 
 \begin{pmatrix}
  x_{hor} \\
  y_{hor} \\
  z_{hor}
 \end{pmatrix}
.\end{equation}

\subsection{Trajectory determination}
\label{trajectory}

\subsubsection{Meteor geometry}
\label{meteor_geometry}

The trajectory of the meteor is determined using the 3D unit vectors of the defined points on the meteor line. These vectors are generated for each camera from the known camera orientation. The vectors point to the meteor trail and are defined in the geocentric coordinate system. Since the meteor line is initially defined in the images using a gnomonic projection, the intersection points of the direction vectors with a unit sphere lie on a great circle. A plane is fitted through all the unit vectors from each station by solving the standard plane equation using least-squares 
\begin{equation}\label{eq:plane}
 \vec{n}(\vec{x}{_0}-\vec{x}) = n_{x}x + n_{y}y + n_{z}z + d = 0 
,\end{equation}
where $\vec{x}{_0}$ = 0 is the origin of the geocentric coordinate system, $\vec{x}$ is the direction vector, $\langle n_x,n_y,n_z \rangle$  are the vector components of the normal vector $\vec{n,}$ and \textit{d} is the distance from the plane to the origin and in this equation is equal to zero. The apparent radiant $RA_{app}$, $Dec_{app}$ of the meteor is calculated as the cross-product of the two normal vectors $\vec{n}{_1}$ $\times$ $\vec{n}{_2}$, determined in (\ref{eq:plane}) with the subscripts indicating the two camera stations. The mean altitude of the meteor is computed by intersecting the direction vector of the central point of the meteor from the shuttered meteor station with the plane generated from the direction vectors to the meteor trail from the second station.

\begin{figure*}
  \centering
  \includegraphics[width=0.45\textwidth]{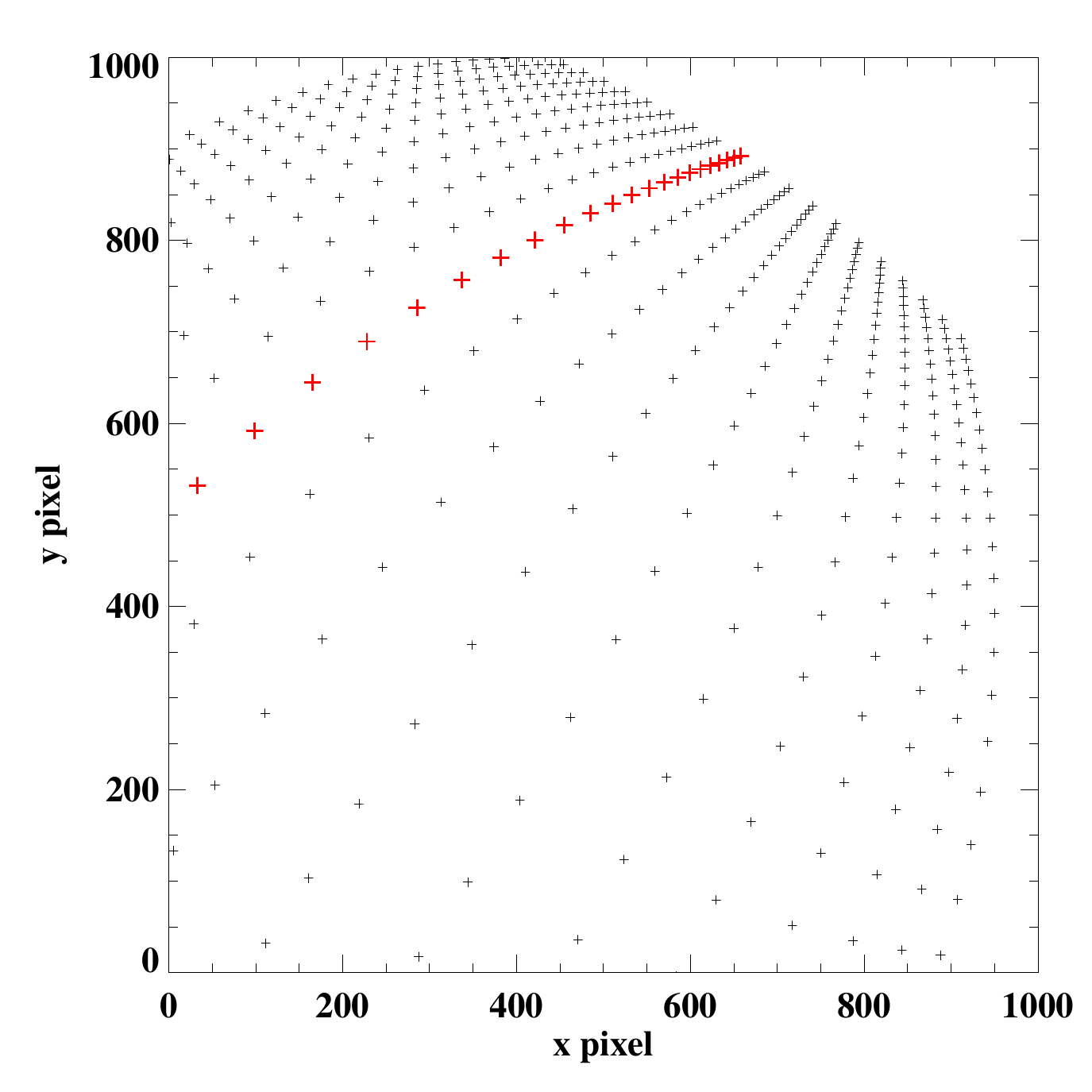}\includegraphics[width=0.45\textwidth]{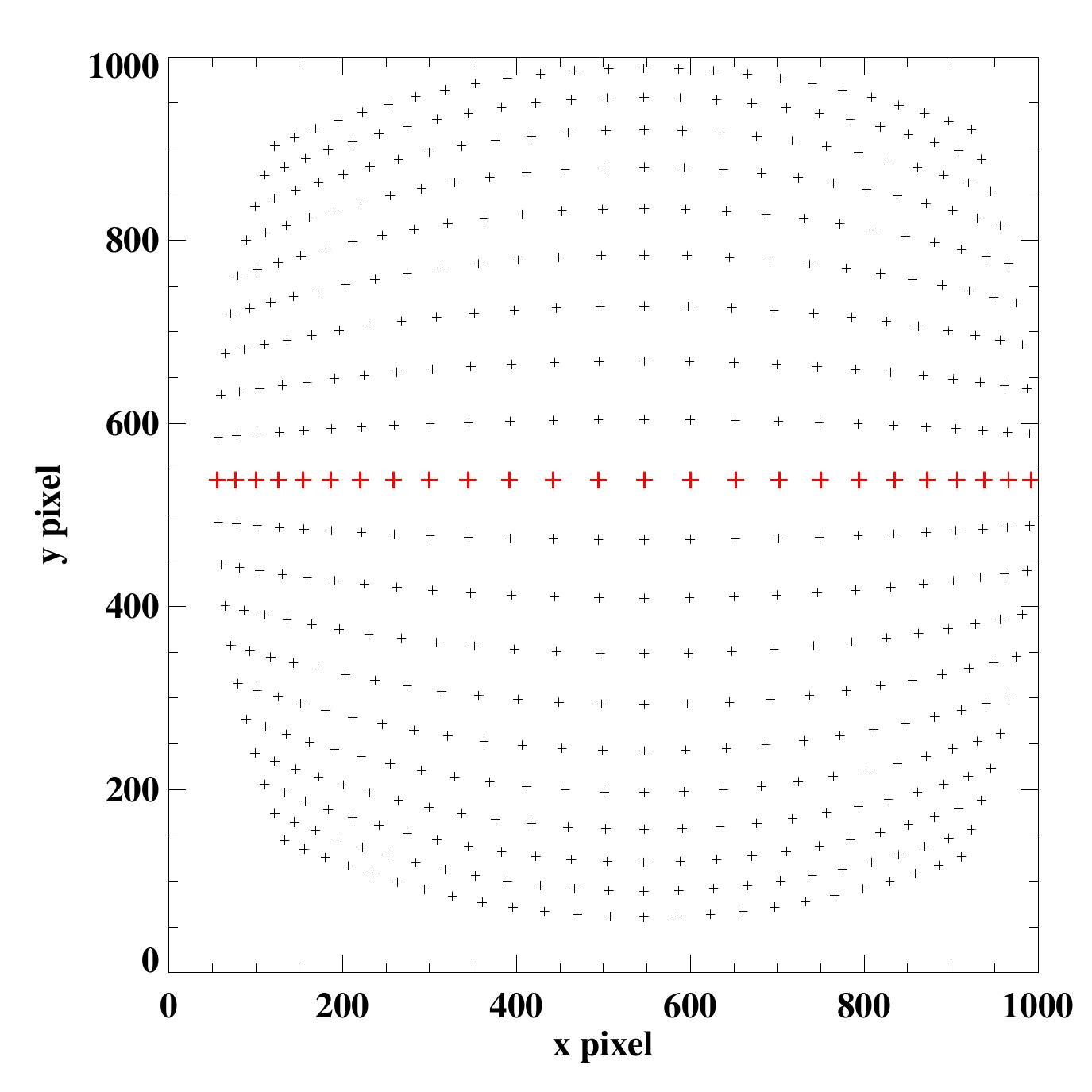}
  \caption{\textit{Left panel}: Reconstructed meteor plane using the determined meteor orientation and position in camera coordinate system. \textit{Right panel}: Normalized meteor plane being parallel to the image plane and at 100 km distance. The crosses in red color highlight the meteor line on both planes.}
  \label{fig:normalized_plane}
\end{figure*}

To determine the speed and duration of a meteor, each shuttered meteor image is compared with a database of synthetic meteors (see Section \ref{validation}). These meteors have a fixed geometry and orientation with respect to the camera, i.e. the meteor is moving parallel to the x-axis of the camera system and at 100 km above the camera. The projection of the meteor position at time interval t=dt/2 coincides with the principal point of the camera. The database is created by varying two parameters: the speed and duration. The step size of the database is 0.1 km s$^{-1}$ for the velocity and 0.02 s for the duration of the meteor. From the known geometric relation between the image and meteor plane, the meteor image is transformed so that the meteor plane becomes parallel to the image plane and the distance between principal point and plane is adjusted to 100 km (Fig. \ref{fig:normalized_plane}). This normalised image is then compared with synthetic meteor images in the database accounting for $(n_s/0.1)\times(n_d/0.02)$ different combinations for speed and duration, where $n_s$ and $n_d$ are the resolution of our partitioning in speed and duration, respectively. For each combination, the meteor trail is time-shifted by 0.06 s to account for various beginning points. The Pearson correlation coefficient is calculated between a synthetic meteor in the database and the normalised image. For the best match we follow a top-down searching approach: first a coarse search is made and then gradually the step size is decreased around the parameters showing a higher correlation. The speed (\textit{V$_{obs}$}) and duration of the meteor are derived from the synthetic image with the highest correlation.

A meteoroid experiences a deceleration when it reaches the denser layers of the Earth's atmosphere. This effect, so-called atmospheric deceleration, depends on the initial speed of the meteoroid and is more prominent for slower meteoroids. In our studies, we are focussing on the fast-moving Perseid meteoroids and therefore, deceleration is ignored here. The Earth's rotational velocity contributes an extra 0.004 $^{\circ}$/s to the calculated right ascension angle of the radiant and is also neglected in this study.

The speed of a meteoroid slightly increases as soon as it experiences the Earth's gravitational attraction, a phenomenon known as zenithal attraction. This effect is computed by performing two integrations following \citet{Jenniskens2011}: one integration backwards in time including the gravitational effects of the Earth-Moon system until the meteoroid reaches the Earth's sphere of influence and a second integration forwards accounting only for the masses of the Sun and the planets. As input for both integrations the state vector of the meteoroid is used. The new state vector yields the position and velocity of a meteoroid at the time it was recorded in the absence of the Earth-Moon system. The velocity vector now points to the geocentric radiant ($RA_{geo}$, $Dec_{geo}$).

\subsection{Heliocentric orbit}
\label{orbit}
The orbital path of a meteoroid around the Sun, requiring knowledge of Earth and Sun positions, is computed using standard solar system ephemerides (DE-421). We use the SPICE software library \citep{Acton2011} to access ephemeris data and retrieve the following geometric transformations.
First the state vectors are transformed from an Earth-centred to a Sun-centred ecliptic coordinate system in J2000, i.e.
\begin{equation}\label{eq:r_eclip}
\vec{r}_{eclip} = \vec{r}_{geo} \cdot R_{geo2eclip}
.\end{equation}
The heliocentric position and velocity vector are simply computed as the vector sums%
\begin{equation}
 \vec{r}_{hel} = \vec{r}_{met} + \vec{r}_{earth}
,\end{equation}
where \textit{r$_{met}$} and \textit{r$_{earth}$} are the state vectors of the meteoroid and the Earth with respect to the Sun in the heliocentric ecliptic coordinate system. Finally, the osculating elements of the orbit are determined using the heliocentric state vector from SPICE routines.

\subsection{Photometric reduction}
\label{photometry}

\subsubsection{Meteor photometry}
\label{meteor_photometry}

Photometric information on meteors is extracted by deconvolving the emitted light of a meteor from the registered signal in equal time intervals \citep{Christou2015}. We remove the effects of radial distortions in the raw image and resample it using inverse distance weighting interpolation. The displacement values $\Delta$$x_{ij}$ and $\Delta$$y_{ij}$ for each pixel are determined from the geometric camera calibration (Section \ref{calibration}). To speed up the interpolation process, we limit the interpolation to a rectangular area around the meteor trail, for which the position is defined (see Fig. \ref{fig:pixonline}). The change of the angular velocity of a meteor owing to perspective distortion is taken into account by projecting the previously determined 3D meteor path (Section \ref{image 2 space}) to the image. 

The number of time intervals $n_{t}$ for which the brightness of the meteor is estimated is computed as the ratio of the length of the rectangular area to the spatial sampling resolution defined by the user. A constant spatial sampling size ensures a stable numerical solution for meteors with low angular velocities, but at the same a high-resolution photometric profile for meteor with high angular velocities. From the estimated meteor velocity, the time the meteor needs to cross the rectangular area is calculated following an iterative process. The photometric model can now be applied to the meteor line in the interpolated image.

\subsubsection{Photometric calibration}

For photometric calibration we use stars depicted in the image. Their positions in the image (pixel coordinates) are computed using the DAOPHOT routines \citep{Stetson1987} and transformed to the equatorial coordinate system at J2000. The stars are identified by querying the VIZIER database \citep{Vizier1997} and matching them to the brightest stars ($m<8$) found within a radius of 30 arcminutes ($\sim$4 pixel) from their position. The flux of each star is measured by defining three circular areas around the light source: an inner circular area measuring the light coming from the star and an outer ring determined by two circular areas defining the sky background. We set the star aperture to a radius of 3$\times$FWHM, which encloses nearly 100\% of the stellar flux \citep{Merline1995}. The instrumental magnitude $m_{inst}$ is then defined as
\begin{equation}
 m_{inst}=A-2.5log_{10}\left(\frac{(\sum_{i=1}^nC_i)-nC_{sky}}{t}\right)
,\end{equation}
where $A$ is an arbitrary constant, $C_i$ is the DN value in the $i^{th}$ pixel, $C_{sky}$ is the mean sky background value, $n$ is the number of pixels in each aperture, and $t$ is the integration time of the frames.

To transform the computed instrumental magnitudes to a standard photometric system, we first convert the Hipparchos $H_p$ magnitudes from the Vizier database into Johnson $V$ magnitudes using the following expression \citep{Harmanec1998}:
\begin{equation}
 V=H_p+a_1(B-V)+a_2(B-V)^2+a_3(B-V)^3+a_4
,\end{equation}
where $B-V$ is the colour index of each star from the VIZIER database and $\alpha_{i}$ the transformation coefficients. The light emitted by each star is partially absorbed by the Earth's atmosphere. Therefore, the amount of the absorption for each star is proportional to the amount of atmosphere the light has to traverse to reach an observer on the Earth's surface. This means that light of stars appearing close to the horizon experiences a greater absorption than stars close to the zenith. The amount of atmosphere, called airmass, is calculated as
\begin{align}
&X=sec(z)-0.00186(sec(z)-1) \\ 
&-0.002875(sec(z)-1)^2 - 0.0008083(sec(z)-1)^3 \nonumber,
\end{align}
where $z$ is the zenith angle \citep{Warner2006}. The equation is taking into account the curvature of the Earth. Moreover, the attenuation of light is computed as a function of the wavelength due to Rayleigh scattering and therefore, the amount of attenuation for each star depends on its colour. To account for the colour difference of our star field, we apply a colour correction for each star using the colour indices from the star catalogue. The instrumental magnitudes are corrected for atmospheric effects and converted to absolute magnitudes,
\begin{equation}
 m_{calib}=m_{inst}+T_{c}CI-Xk-Z_{pI}
,\end{equation}
where $X$ is the airmass, $T_c$ the transformation coefficient, $CI$ the colour index, $k$ is the extinction coefficient given in magnitudes per unit airmass, and $Z_p$ is a scaling factor. We compute three sets of correction parameters for $U,V,$ and $I$ colour corrections using least squares. 

\begin{figure}\centering
  \includegraphics[width=0.5\textwidth]{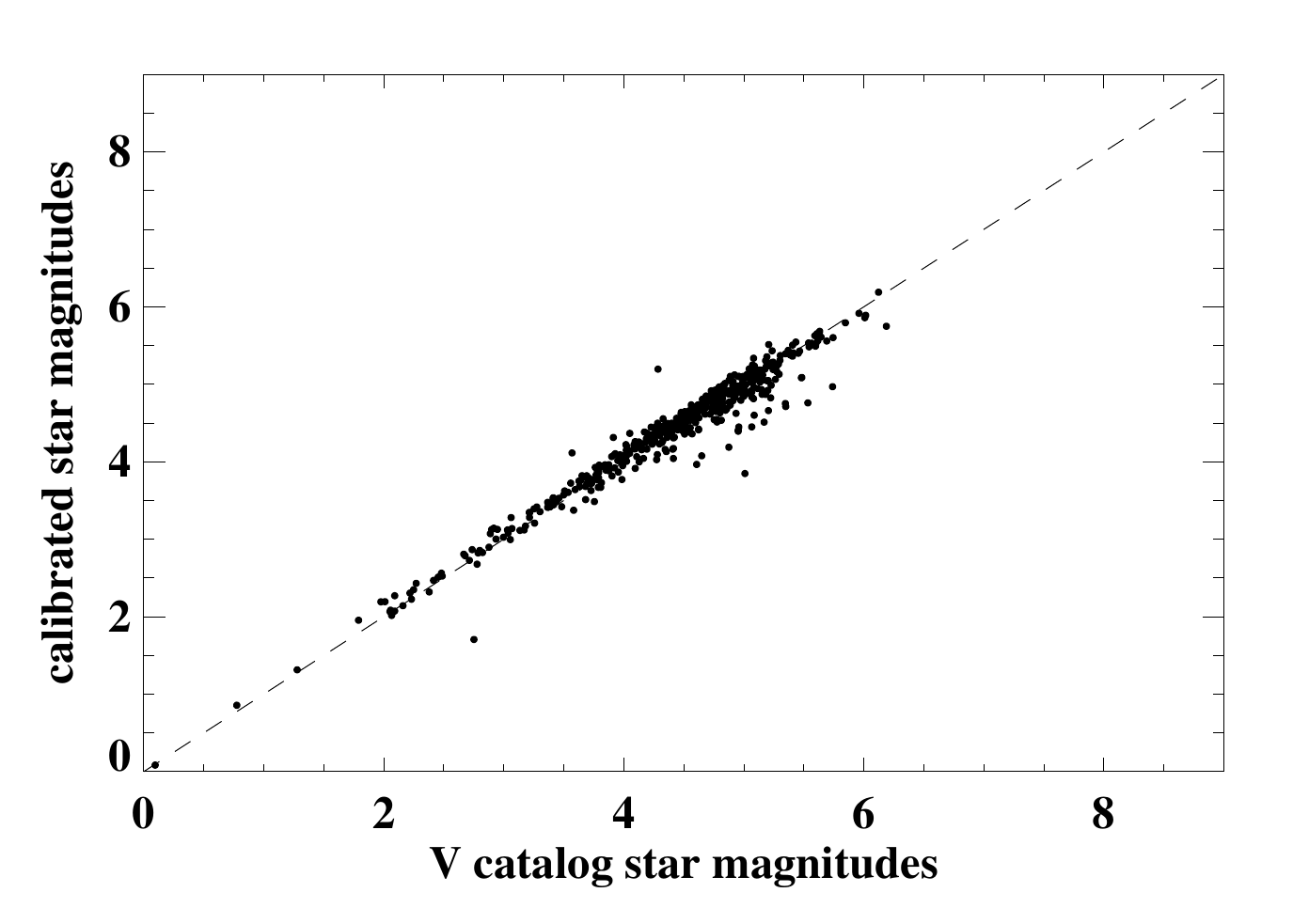}
  \caption{Calibrated star magnitudes vs. standard star V magnitudes. Dashed line shows the ideal one-to-one relationship between the two quantities.}
  \label{fig:vmag}
\end{figure}

We tested our photometric calibration module with a typical SPOSH image. We detected 393 stars in the image, where the faintest is of +6.3 magnitude. A high correlation between calibrated and standard stellar magnitudes using the $V-I$ colour index was found, matching the spectral response of the system. Stars with an airmass greater than 3 were excluded from the procedure. Figure \ref{fig:vmag} shows the calibrated magnitudes of the stars with respect to their catalogue magnitudes from a single frame. The standard deviation between catalogue and measured magnitudes was of 0.22 magnitudes.

\section{Error propagation}
\label{error_propagation}

The errors of the unknown parameters are calculated by applying error propagation. In the sections to follow we refer to these as the propagated uncertainty (or propagated errors) to distinguish  this uncertainty from the statistics of differences between estimated and a priori known parameters (estimated uncertainty) as well as the uncertainty in estimating a common property of the meteors, for example the radiant and speed of a shower, by taking the average over a number of meteors (observed uncertainty). We encounter the estimated uncertainty principally in tests with our synthetic data (Section \ref{synthetic_data}). The unknown parameters are the apparent and geocentric radiant positions, observed, geocentric, and heliocentric speed of the meteoroid, and orbital elements of its orbit around the Sun. The observed quantities are the parameters of the meteor line $\rho$ and $\theta$. The general law of error propagation is of the form
\begin{equation}
 C_{yy}=\frac{\partial y}{\partial x}C_{xx}\frac{\partial y}{\partial x}^{T}
,\end{equation}
where $C_{xx}$ is the stochastic model of the measurements and $y$ are the parameters to be estimated.  The parameter $C_{yy}$ is the variance-covariance matrix of the unknown parameters. The uncertainties of the direction and location of the meteor line on the image affect the uncertainties of the parameters and need to be carefully estimated. We used our synthetic meteor dataset (see Section \ref{validation}) to estimate the line uncertainties. The distribution of the estimated uncertainties for $\rho$ and $\theta$ are well approximated by a normal distribution with a standard deviation of 0.07$^{\circ}$ for $\theta$ and 1.35 pixel for the distance of the projection to the line. These values depend highly on the length of the meteor trail, PSF, and resolution of the CCD. 

\section{Software validation}
\label{validation}

\subsection{Synthetic meteor data}
\label{synthetic_data}

We verified our software modules with the help of synthetic meteor data. A meteor trail is generated by providing a number of parameters which i) define the dynamic and photometric properties of a meteor, ii) define the geometric relation between the observers and the meteor, and iii) projects the luminous path of the meteor to the image plane of the given camera system. Table \ref{table:parameters} summarises the initial conditions used to generate synthetic meteor trails. 

Once the meteor path is generated in space, the corresponding meteor trail is projected on the image plane. The trail is created by convolving a 2D Gaussian curve imitating the motion of a point-like light source on a given camera system. The method is based on the photometric model in \citet{Christou2015} implemented in reverse. The peak intensity value of each meteor is kept constant while the standard deviation of the Gaussian PSF is set equal to one pixel. The brightness of each synthetic meteor is normally distributed along the meteor trail. The peak brightness also varies between each meteor and resembles different shape curves \citep{Beech2004,Borovicka2007}. The position of the peak along the meteor trail in our sample follows a normal distribution with its centre at \textit{n}$_{t}$/2 and a standard deviation of \textit{n}$_{t}$/10, where \textit{n}$_{t}$ is  the number of time intervals. The meteor trails in one of the stations were chopped periodically to simulate the effect of the rotating shutter. The starting point of a meteor at time \textit{$t_0$} is placed randomly within a shutter break and ranges between 0 s and 0.06 s. As an example, a meteor with \textit{$t_0$}=0 will receive light directly, while a meteor with \textit{$t_0$}=0.06 occurs exactly at the time when the shutter is located in front of the lens. For the parameters of the inner orientation of the camera, typical values for the SPOSH camera were used. The pointing of both cameras was chosen to slightly deviate from an optical axis parallel to the zenith. The distance between the two stations was set at 55.6 km. All synthetic meteors in our simulations occur at the same time, i.e. time information is not relevant. The position of the radiant is given in equatorial coordinates system as RA and Dec.

\begin{table}
 \centering
 \small
 \caption{Initial conditions for synthetic meteors with random radiant positions discussed in Section \ref{synthetic_data}. The camera parameters are typical values for the SPOSH camera.}
 \begin{tabular}{llll}  
  \toprule
  \multicolumn{4}{l}{\textbf{Meteor parameters}} \\
        \midrule

        \textbf{position} & lat ($^{\circ}$) & lon ($^{\circ}$) & alt   (km)    \\
        meteor trail      & -2.8 -- 2.8 & -2.3 -- 2.3   & 75--125 \\
        \textbf{direction}& azimuth ($^{\circ}$) & elevation ($^{\circ}$)       \\
        velocity vector   & 0-360  & 30-60      \\
        speed             & \multicolumn{3}{l}{20-75 km s$^{-1}$}       \\
        duration          & 0.2-0.6 sec                         \\
        time resolution   & n$_{t}$=30                          \\

        \midrule
  \multicolumn{4}{l}{\textbf{Camera parameters}} \\
        \midrule

        \textbf{position} & lat ($^{\circ}$) & lon ($^{\circ}$) & alt   (km)    \\
        camera$_{1}$      & 0.25 & 0.0 & 0.0         \\
        camera$_{2}$      & $-$0.25 & 0.0 & 0.0         \\
        \multicolumn{4}{l}{\textbf{orientation}} \\
        interior param.   & x$_{p}$=519.5 pix& y$_{p}$=513.5 pix& c=7 mm        \\
        exterior param.   & $\omega$= -1.0$^{\circ}$ & $\phi$= -3.4$^{\circ}$ & $\kappa$= 2.2$^{\circ}$       \\
        \multicolumn{4}{l}{\textbf{CCD}} \\  
        sensor size       & 1024 pix & 1024 pix         \\
        pixel size        & 13 $\mu$m  & 13 $\mu$m              \\

  \bottomrule
  \end{tabular}
\label{table:parameters}
\end{table}

An image with random noise was generated using the noise properties of the SPOSH images that are the same size as the meteor image. Additionally, 20 2D-Gaussian PSF simulating the stellar sources in the image were distributed randomly in the FOV of the camera. The image with the synthetic stars was added to the noise image. The position and brightness levels of the stars were kept fixed for all synthetic images. The light of the synthetic stars in the noise image represent the remaining light due to scintillations, visible in the SPOSH difference image data. Finally, the noise image was added to the meteor image, creating the input for our algorithm. The software uses the images of each synthetic meteor as input and calculates its radiant position, speed, brightness, and heliocentric orbit.

\begin{figure*}
  \centering
  \includegraphics[width=0.5\textwidth]{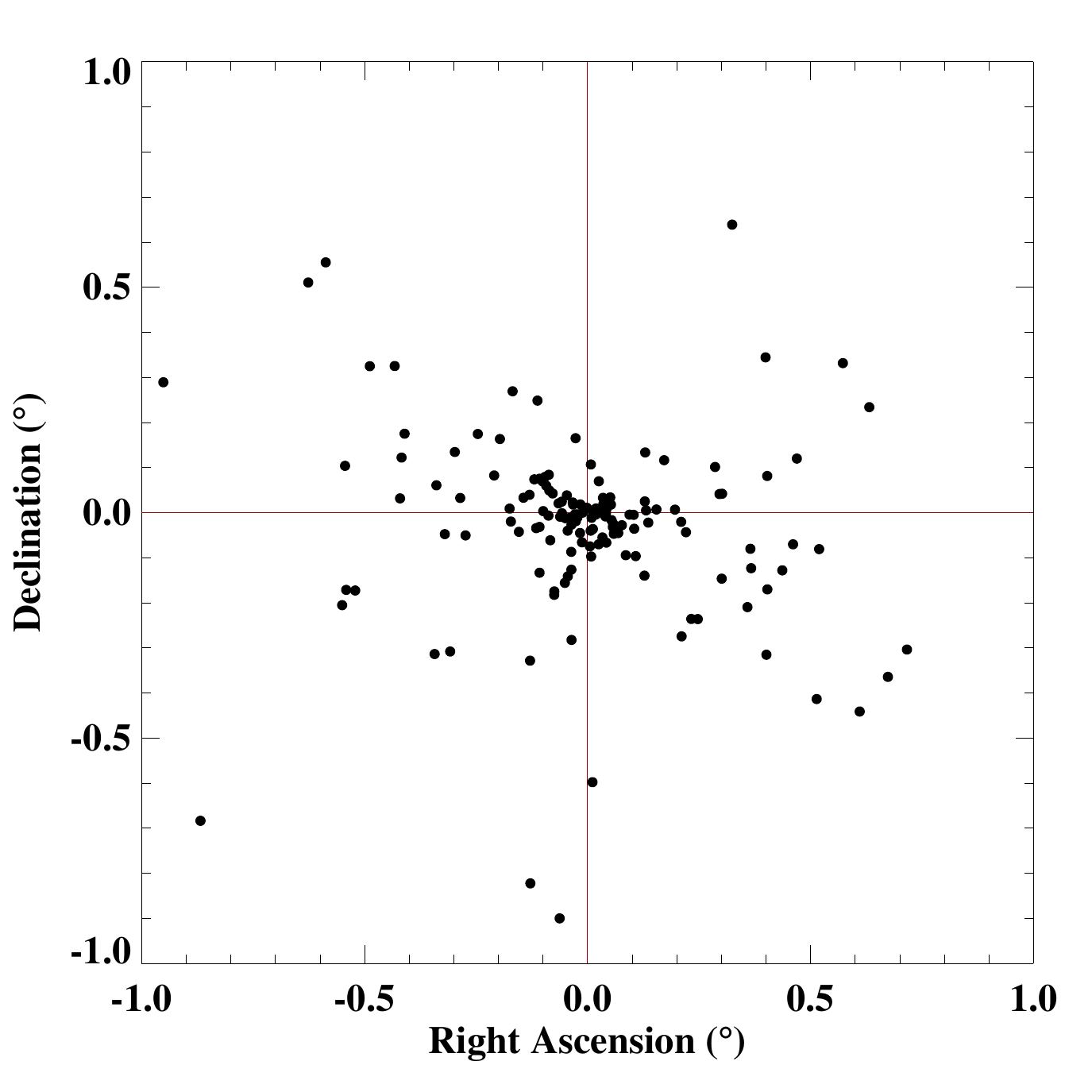}\includegraphics[width=0.5\textwidth]{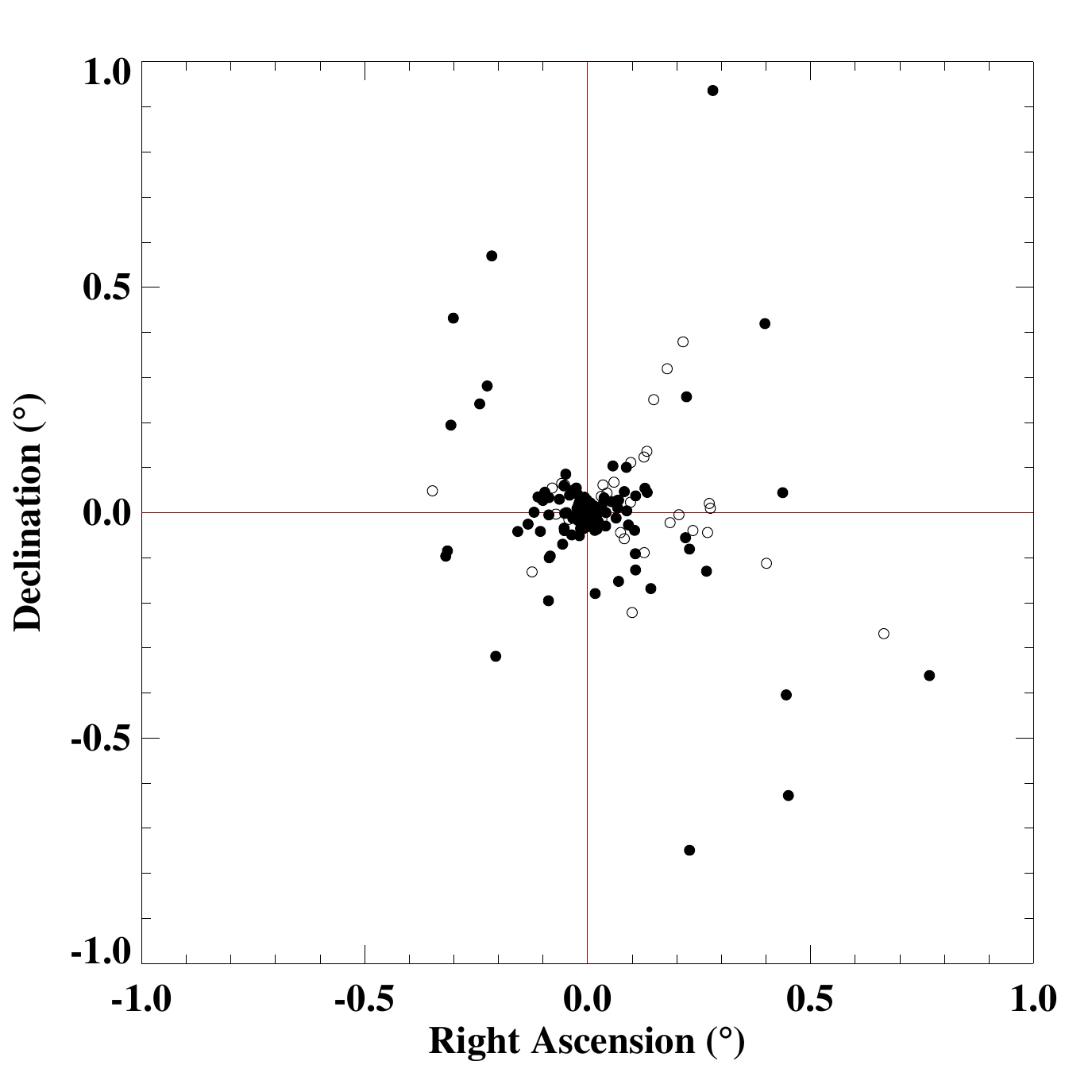}
  \caption{\textit{Left panel:} Radiant dispersion for synthetic meteors originating from random directions. \textit{Right panel:} Radiant dispersion for meteors with the radiant point located in the local zenith. The filled circles ($\bullet$) represent meteors appearing 50$^{\circ}$ above the horizon while open circles ($\circ$) show meteors with elevation angles lower than 50$^{\circ}$.}
  \label{fig:radiant_spread}
\end{figure*}

\begin{figure}\centering
  \includegraphics[width=0.5\textwidth]{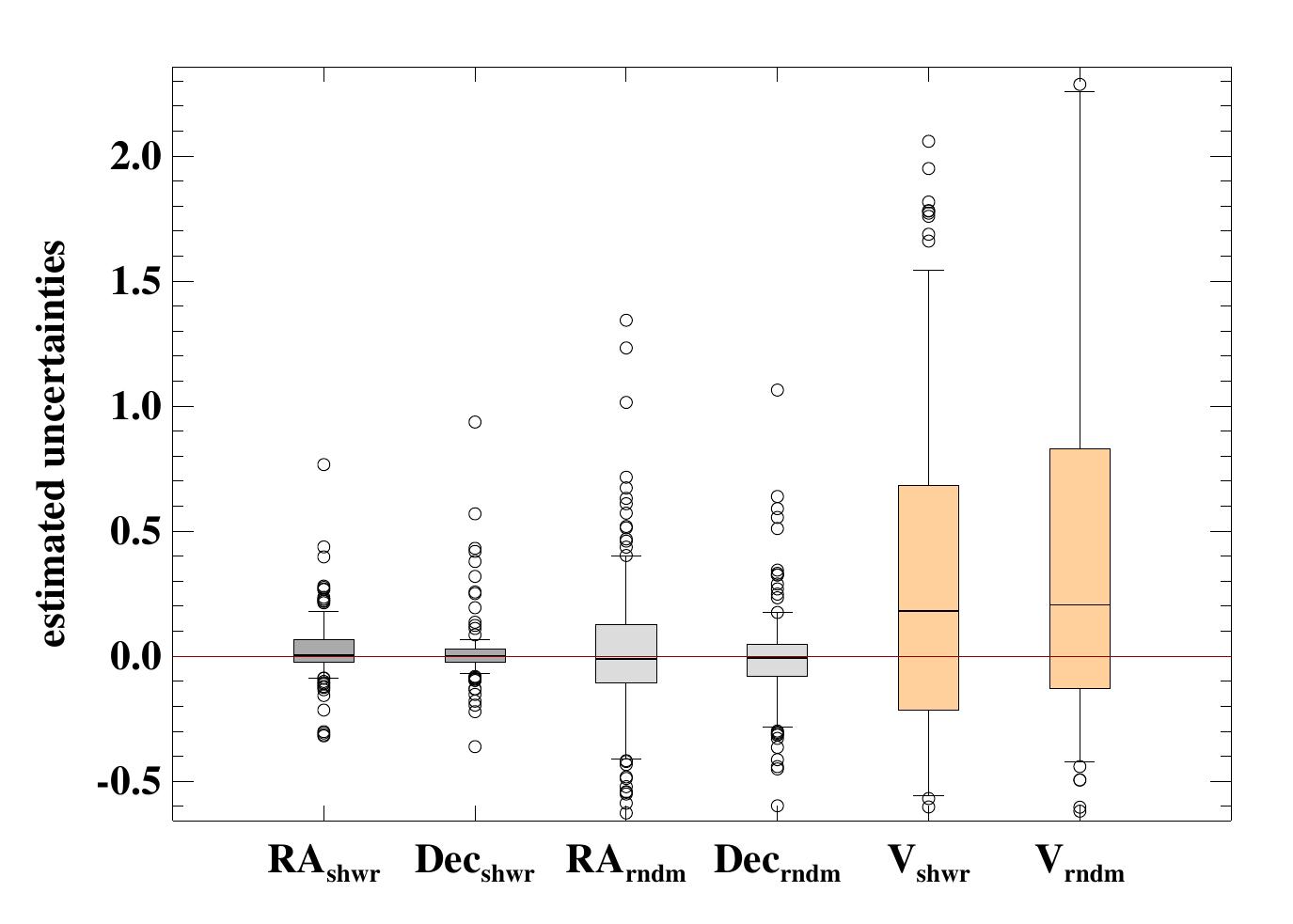}
  \caption{Distribution of the estimated uncertainties in RA, Dec and speed for the synthetic meteors. The length of the boxes indicates the dispersion of the data. Each box encloses 50\% of the data. The extending vertical lines from the boxes indicate the range of 80\% of the data with the lower and upper horizontal bars marking the 10\% and 90\% levels. Data outside the 80\% range are shown as open circles ($\circ$). The horizontal line inside each box indicates the median value. The units are degrees for RA and Dec, and km $s^{-1}$ for speed.}
  \label{fig:box_res}
\end{figure}

\begin{figure}\centering
  \includegraphics[width=0.5\textwidth]{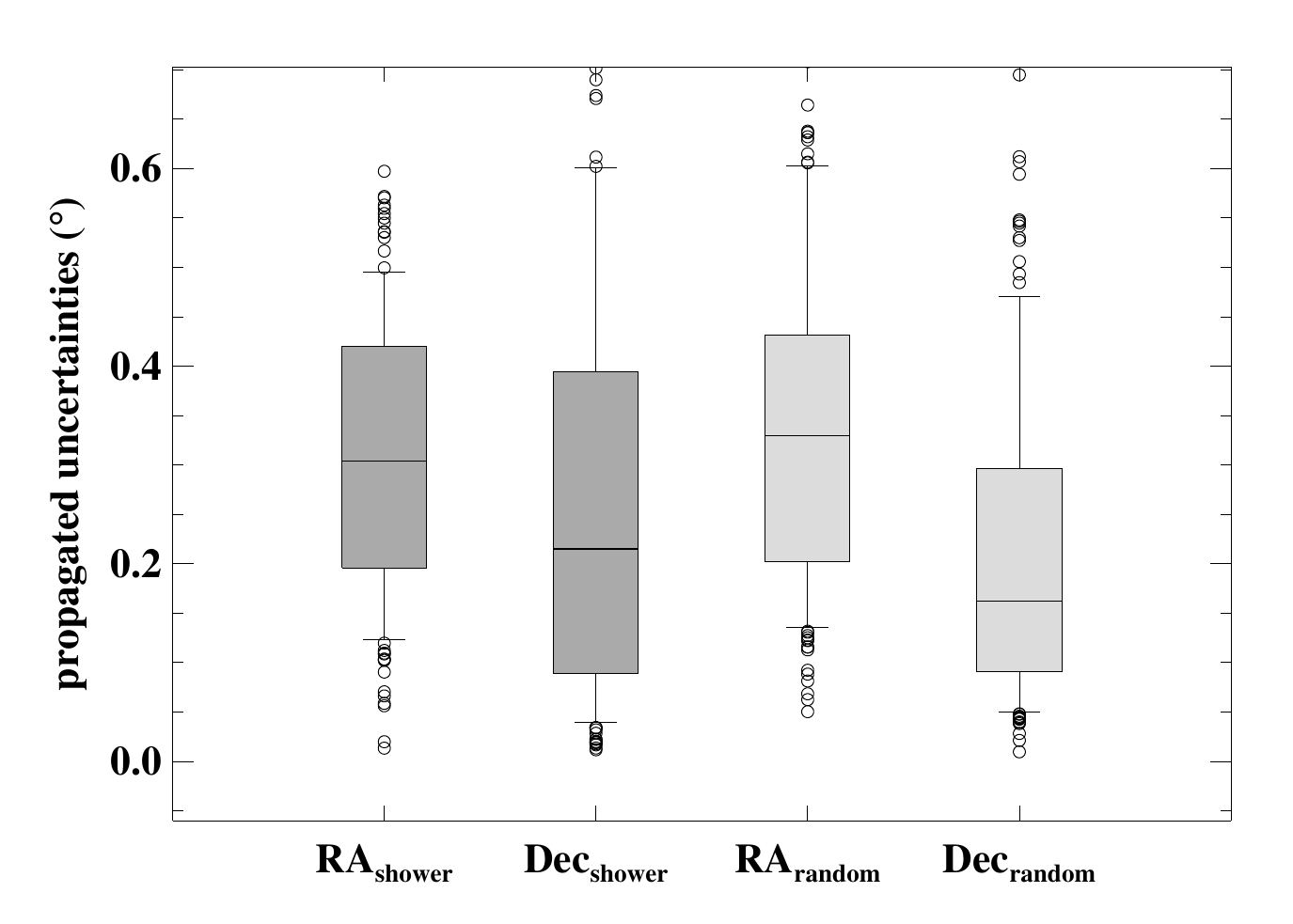}
  \caption{Distribution of propagated uncertainties for the synthetic meteors. For a description of the plot see the caption of Figure \ref{fig:box_res}.}
  \label{fig:box_std}
\end{figure}

We generated a dataset of synthetic meteor trails considering different geometric configurations. We present results for two types of synthetic data. For the first, we created 208 synthetic meteors with random positions and directions with respect to the location of the cameras, and then used our program to estimated their radiants, speeds, and magnitudes. Forty-seven\ of the synthetic meteors had convergence angles $Q \leq$ 10$^{\circ}$ yielding large errors in the radiant determination. These meteors were therefore excluded from the procedure. For the remaining 161 synthetic meteor pairs with $Q >$ 10$^{\circ}$ we determined the radiant position (RA and Dec) and the speed (V) and computed their estimated uncertainties as the difference between the a priori value and that calculated from the code. We describe the statistical dispersion of the probability distributions by calculating the median absolute deviation (MAD), which is statistically a more robust measure for asymmetric distributions than the standard deviation. The distributions of the estimated uncertainties for RA and Dec were centred at zero with a median deviation of 0.11$^{\circ}$ and 0.07$^{\circ}$, respectively (Fig. \ref{fig:radiant_spread}). The statistical properties of the estimated uncertainties are shown graphically in Figure \ref{fig:box_res}. The propagated uncertainties had a median value of 0.33$^{\circ}$ for RA and 0.16$^{\circ}$ for Dec (Fig. \ref{fig:box_std}). The propagated and estimated errors are in good agreement, which implies a realistic stochastic model (Table \ref{table:errors}). The distribution of the estimated errors for the speed appears to be offset from zero with a median of 0.24 km $s^{-1}$ and a median deviation of 0.51 km $s^{-1}$. For calculating the meteor magnitudes, we used a subset of the data consisting of 185 un-shuttered meteors of which the individual residual is $\leq$ 0.5 m. The estimated uncertainties had a median value of +0.01 m and a median deviation of 0.03 m.

\begin{figure}\centering
  \includegraphics[width=0.5\textwidth]{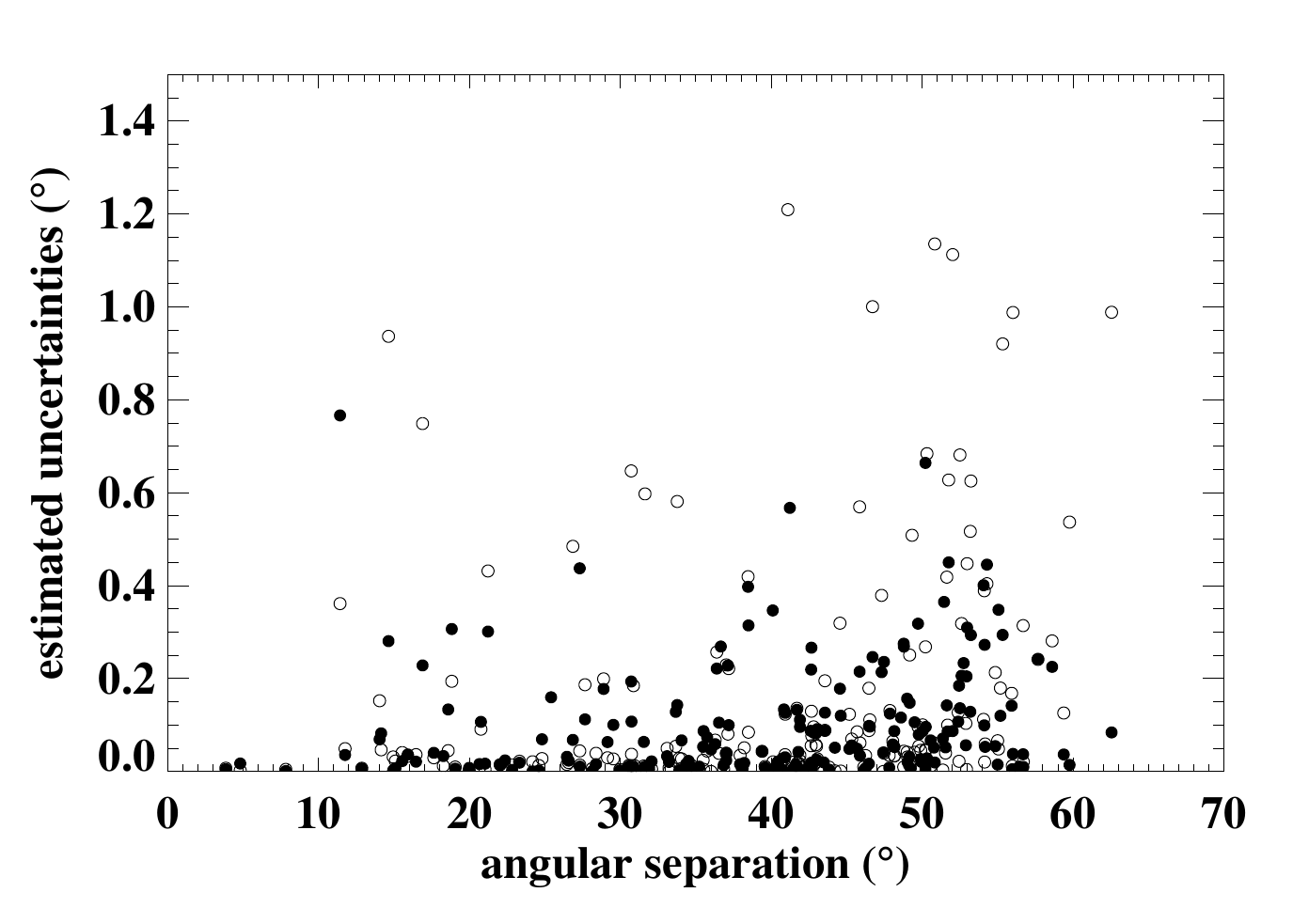}
  \caption{Plot showing the relation between the angular separation and the estimated uncertainties in right ascension (filled circles) and declination (open circles).}
  \label{fig:angDist}
\end{figure}

A second set of 208 synthetic meteors was then created with the same radiant point for all meteors placed at the local zenith to simulate a meteor shower observed by the two cameras. One hundred seventy of the meteors had a convergence angle $Q >$ 10$^{\circ}$. Nine radiants with large estimated uncertainties were excluded from the procedure. As for the first set of synthetic data, the estimated errors for RA and Dec also have a zero median but slightly lower median deviation of 0.05$^{\circ}$ and 0.03$^{\circ}$ , respectively (Fig. \ref{fig:box_res}). These reduce to 0.03$^{\circ}$ and 0.03$^{\circ}$ when considering only 126 meteors occurring $>$ 40$^{\circ}$ from the local horizon. The dispersion of the propagated uncertainties for RA and Dec are similar to those computed for the first synthetic dataset. The median propagated uncertainty was 0.37$^{\circ}$ for RA and 0.27$^{\circ}$ for Dec (Fig. \ref{fig:box_std}). Figure \ref{fig:angDist} shows the relation between the position of a meteor in the image and the estimated errors in radiant. Since the pointing of the camera and the radiant of the simulated shower was set to the local zenith, the angular separation between the radiant and position of a meteor in the image corresponds to the elevation angle of the meteor. The median for the estimated errors in the speed for these 161 meteors was 0.12 km $s^{-1}$. The median of the estimated errors for 192 un-shuttered synthetic meteors, for which the residuals in magnitude \textit{$\Delta$mag} were < 0.5m, was +0.01 m with a median deviation of 0.02 m. 

\begin{table}
 \centering
 \small
\begin{threeparttable}
 \caption{Different types of uncertainties for the synthetic and real meteor data.}
 \begin{tabularx}{0.45\textwidth}{c|l|c|c|c}
  \toprule
        \multicolumn{5}{l}{ \textbf{Synthetic meteor data} } \\
        \midrule
        \multicolumn{2}{l}{} & \multicolumn{3}{c}{uncertainties} \\
        \midrule
        N & & \multicolumn{1}{p{1.5cm}|}{MAD\tnote{1} \enspace \tiny{(estimated)}} & \multicolumn{1}{p{1.5cm}|}{median \enspace \tiny{(propagated)}} & \multicolumn{1}{p{1.5cm}}{MAD\tnote{1} \enspace \tiny{(propagated)}} \\
        \hline  
        
        151      & RA & 0.11 & 0.33 & 0.12  \\
        (random) & Dec & 0.07 & 0.16 & 0.09  \\
                 & V & 0.40 & \textellipsis & \\ 
        \hline
        161      & RA & 0.05 & 0.35  & 0.14    \\
        (shower) & Dec & 0.03 & 0.25  & 0.16    \\
                 & V &  0.45   &  \textellipsis & \textellipsis \\      
        \hline
         126\tnote{2}    & RA & 0.03 & 0.37 & 0.13 \\
         (shower)        & Dec & 0.03 & 0.27 & 0.18 \\
                   & V  & 0.44 & \textellipsis & \textellipsis \\       
        \midrule
        \multicolumn{5}{l}{\textbf{Real meteor data}} \\
        \midrule

        177 & RA & \textellipsis  & 0.64 & 0.29  \\
        (all) & Dec & \textellipsis      & 0.29 & 0.18 \\
            & V$_{g}$ & \textellipsis      &    1.18 & 0.70 \\
        \hline
        71  & RA & \textellipsis            & 0.72 & 0.21 \\
        (Perseids) & Dec & \textellipsis & 0.22 & 0.22  \\
            & V$_{g}$ & \textellipsis     & 0.88 & 0.48  \\
  \bottomrule
\end{tabularx}
  \begin{tablenotes}\footnotesize
  \item[1] For a symmetric distribution the median absolute deviation equals half the interquartile range. Figure \ref{fig:box_res} shows graphically the statistical properties of the propagated uncertainties.
  \item[2] Meteors with elevations > 40$^{\circ}$.
  \item[] Uncertainties for geocentric radiants in ($^{\circ}$) and for \textit{V} and \textit{V$_{g}$} in (km s$^{-1}$).
  \end{tablenotes}
\label{table:errors}
\end{threeparttable}
\end{table}

\subsection{Real meteor data}
\label{real_data}

We applied our software to three hours of double-station SPOSH image data acquired during an observing campaign held in Greece (i.e. at $\sim$37$^{\circ}$N latitude ) on 12 August 2015 from 23 to 02 UT. The cameras were pointing to the zenith with their x-axis orientated to the north. The baseline between the two sites was 51.5 km. We reduced 177 meteor image pairs and determined their trajectories, velocities and heliocentric orbits. 

\begin{figure}\centering
  \includegraphics[width=0.5\textwidth]{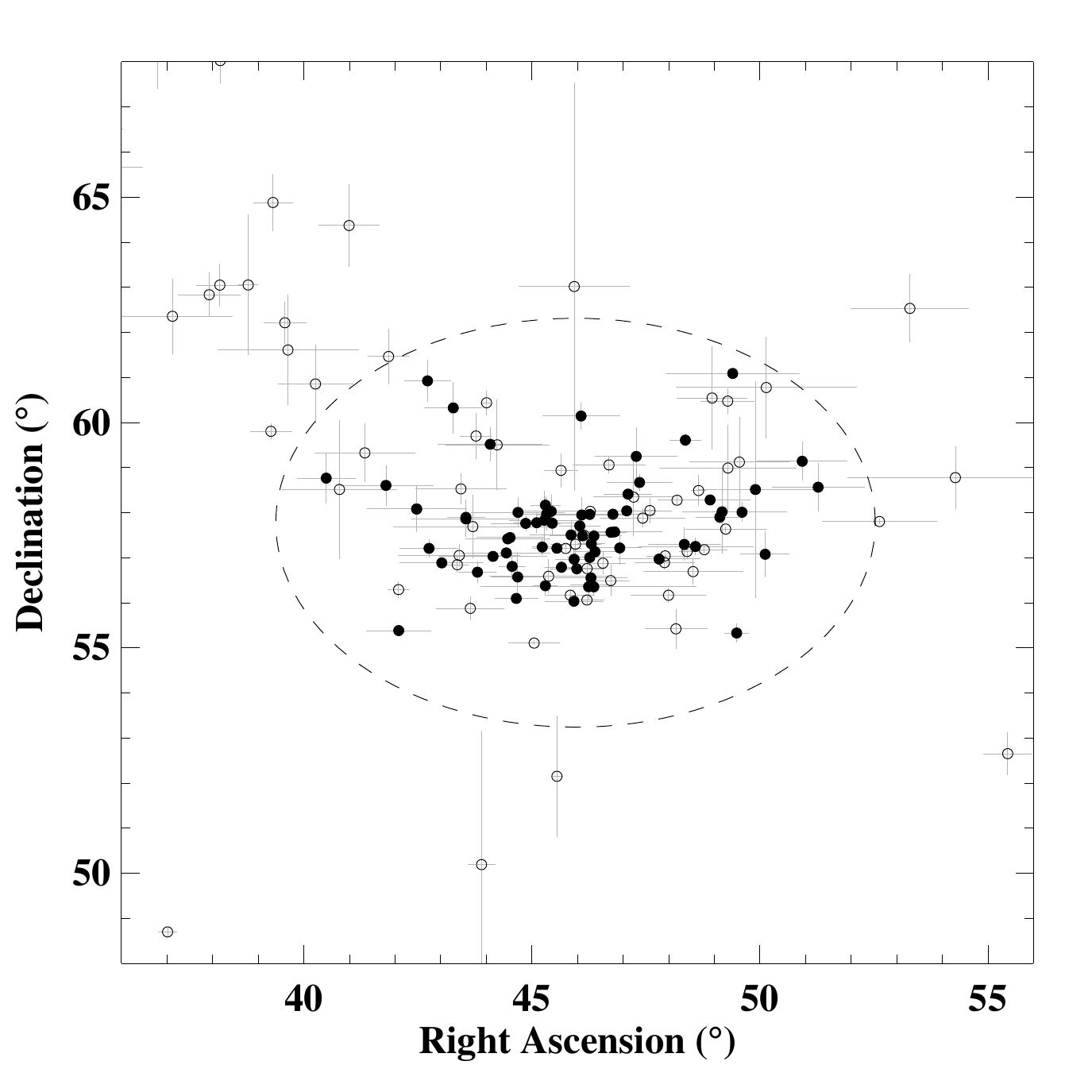}
  \caption{Geocentric radiants of 132 meteors originating close to the Perseid radiant. The ellipse defines the area occupied by Perseids. Meteors classified as Perseids according to their speed are shown as filled circles ($\bullet$). All other meteors not meeting the requirements to be classified as Perseids are shown as open circles (o).}
  \label{fig:radiant_real}
\end{figure}
\begin{figure}\centering
  \includegraphics[width=0.5\textwidth]{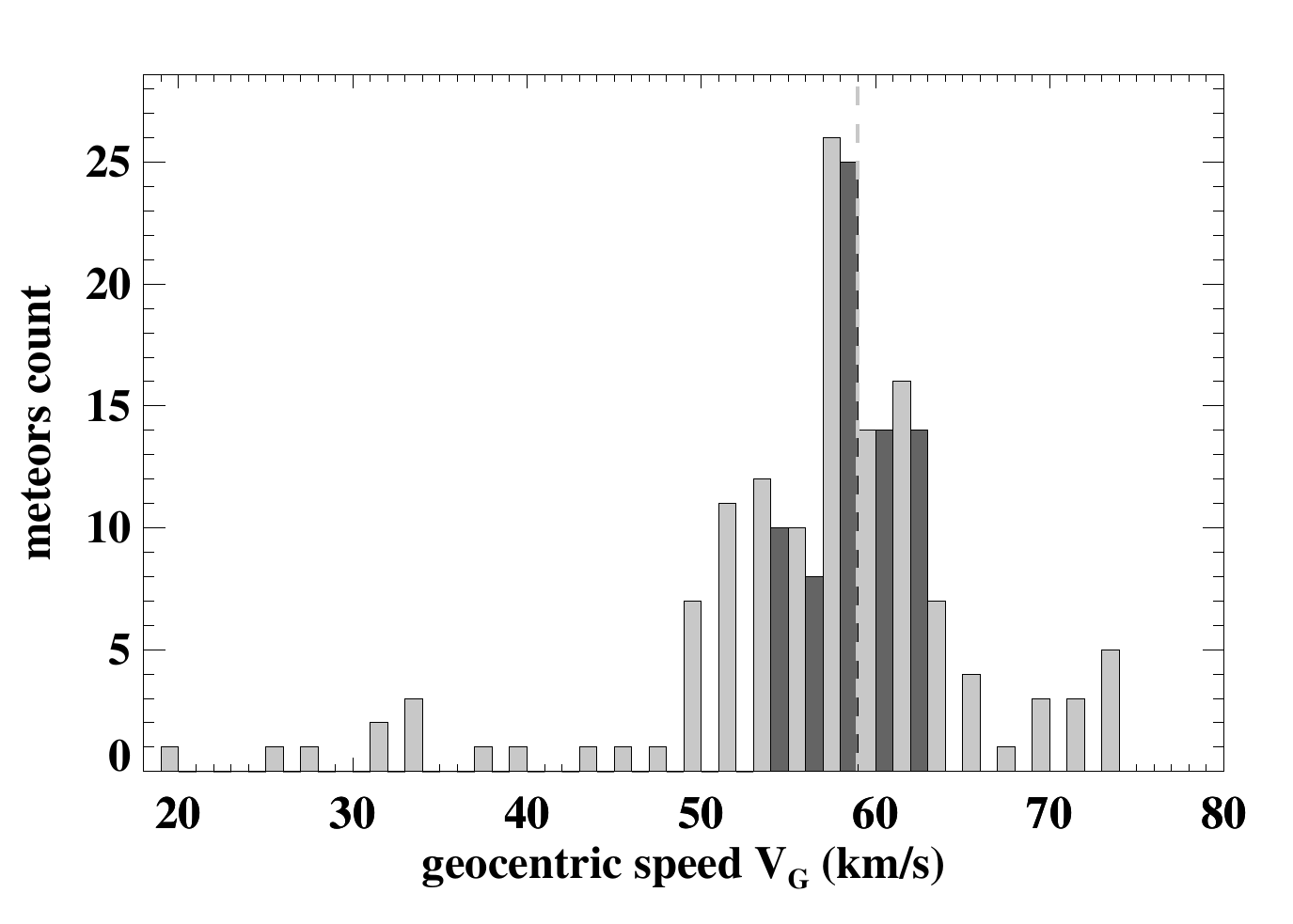}
  \caption{Speed distribution of 132 meteors originating from an area close to the Perseid radiant. The dashed line shows computed median speed of 59.58 km s$^{-1}$. The bars in dark grey show the distribution of speeds for 71 meteors classified as Perseids found within the same area.}
  \label{fig:velDist}
\end{figure}

We focus on a 20$\times$20$^{\circ}$ area centred at RA=46$^{\circ}$, Dec=58$^{\circ}$, close to the nominal radiant position of the Perseids (Fig. \ref{fig:radiant_real}). To distinguish between Perseid and non-Perseid meteors, we performed a classification based on radiant position and speed as follows: 132 meteors were found to radiate from within this area. We determine the radiant of the Perseid shower as the median value of these radiants: RA=45.96$^{\circ}$ and Dec=57.77$^{\circ}$. We assume that most of the meteors are Perseids and we find the 1$\sigma$ uncertainty in RA and Dec to be 3.29$^{\circ}$ and 2.27$^{\circ}$, respectively. The geocentric speeds \textit{V$_G$} have a median value at 58.97 km s$^{-1}$ and a median propagated error of 1.21 km s$^{-1}$ (Fig. \ref{fig:velDist}). We classify all meteors with speeds closer to this median speed than four times the median propagated error, as Perseids. In this way, we identified 71 meteors belonging to the Perseids meteor shower (Fig. \ref{fig:radiant_real}). Their median speed \textit{V$_G$} was found to be 59.58 km s$^{-1}$ with a median deviation of the observed uncertainties of 0.48 km s$^{-1}$. Statistical properties are given in Table \ref{table:errors}. As the aim of our example is to demonstrate the successful usage of our software using real data, we neglected the effect of radiant drift.

\begin{table*}
 \centering
 \small
\begin{threeparttable}
 \caption{Radiant positions, speeds, and orbital elements of Perseid meteors found in 5 studies compared with median values computed in this work and the orbit of the parent comet. Increments of 0.86$^{\circ}$ and 0.51$^{\circ}$ have been added to the median radiant position in RA and Dec to account for radiant shift \citep{Jenniskens2006Book} to the location predicted for 13 August at 7:45 UT.}
 \begin{tabularx}{0.84\textwidth}{l|rrrrrrrrrr}
  \toprule
        & RA & Dec & V$_g$ & a & q & e & i & $\omega$ & Node & N \\
  \midrule
  Jenniskens et al. & 48.2 & +58.1 & 59.1 & 9.57 & 0.949 & 0.950 & 113.1 & 150.4 & 139.3 & 4367 \\
  SonotaCo          & 47.2 & +57.8 & 58.7 & & & & & & & 3524 \\
  Jopek et al.      & 47.3 & +58.2 & 59.0 & \textellipsis & 0.948 & 0.951 & 112.7 & 150.3 & 139.4 & 33 \\
  DMS \tnote{1}     & 48.3 & +58.0 & 59.38 & 71.4 & 0.953 & \textellipsis & 113.22 & 151.3 & 140.19 & 87 \\
  Kres\'{a}k \& Porub\u{c}an & 46.8 & +57.7 & 59.49 & 24.0 & 0.949 & 0.960 & 113.0 & 150.4 & 139.7 & \textellipsis \\ 
  \midrule
  This study      & 46.84 & +58.08 & 59.58 & 2.69 & 0.963 & 0.953 & 113.5 & 153.8 & 139.77 & 71 \\
  (median error)  &  0.72 &  0.21 & 0.88 & 3.09 & 0.01 & 0.06 & 0.8 &  2.8 & 5$\times$10$^{-5}$ \\
  \midrule
  109P (parent comet) & 45.8 & +57.7 & 59.41 & 26.092 & 0.960 & 0.963 & 113.45 & 152.98 & 139.38 \\
  \bottomrule
 \end{tabularx}
 \begin{tablenotes}\footnotesize
 \item[] Orbital elements in epoch J2000; symbols: \textit{a} = semi-major axis (AU), \textit{q} = perihelion distance (AU), \textit{e} = eccentricity), \textit{i} = inclination ($^{\circ}$), \textit{$\omega$} = argument of perihelion ($^{\circ}$), \textit{Node} = ascending node ($^{\circ}$), \textit{N} = number of observed meteors
 \item[1] Dutch Meteor Society 2001: values for the parameters are given in \textit{Meteor Data Center} IAU database (no reference given)

 \end{tablenotes}
\label{table:table2}
\end{threeparttable}
\end{table*}

\begin{figure}\centering
  \includegraphics[width=0.5\textwidth]{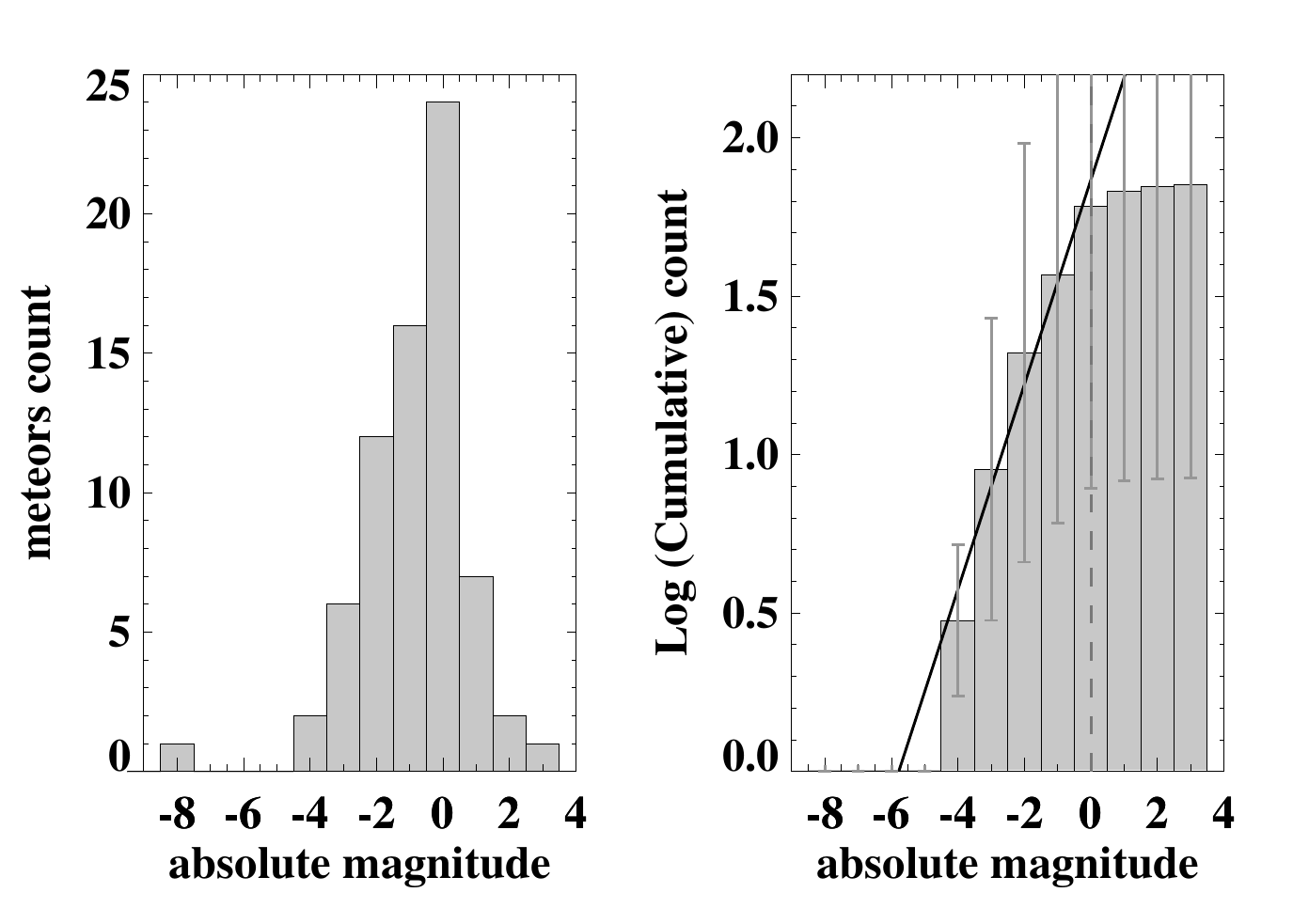}
  \caption{\textit{Left panel}: Magnitude distribution of all meteors located inside the ellipse in Fig. \ref{fig:radiant_real}. \textit{Right panel}: Cumulative distribution of the magnitudes. The slope of the straight line defines the mass distribution index $r$ for the shower during the observing time. Only meteors brighter than +0 m were included in the fit. The dashed line represents the cut-off value.}
  \label{fig:mag_dist}
\end{figure}

\begin{figure}\centering
  \includegraphics[width=0.5\textwidth]{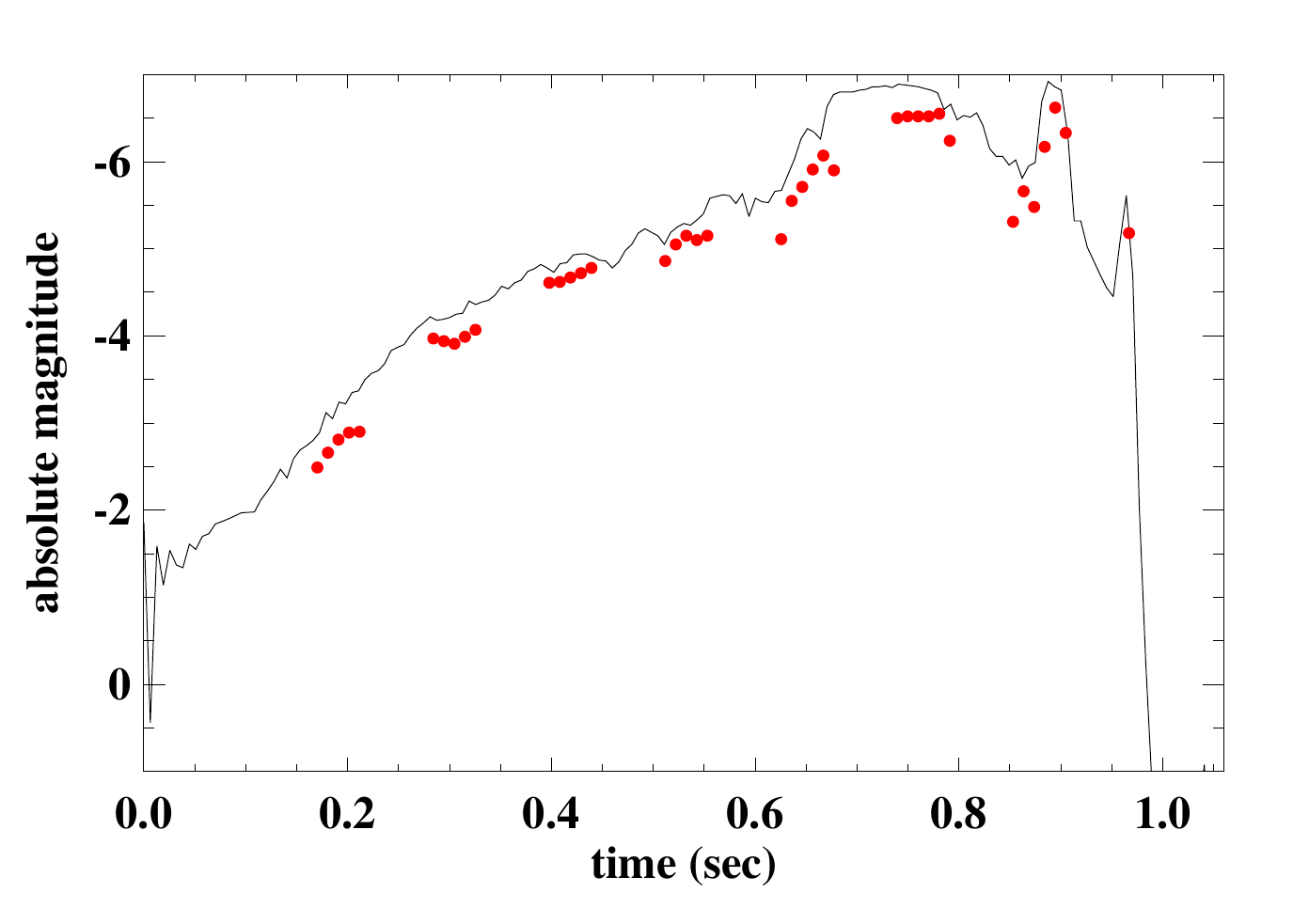}
  \caption{Brightness profile as a function of time for a bright Perseid meteor. The line shows the absolute magnitude calculated from the continuous meteor trail while the red dots represent brightness calculation from the shuttered meteor trail. We note the two flares at $\sim$0.9 s and $\sim$1.0 s.}
  \label{fig:lightcurve}
\end{figure}

We calculated the magnitudes of the 71 meteors identified as Perseids from the un-shuttered images. We defined this magnitude to be the brightest value obtained for the light curve of each meteor. The magnitude distribution index $r$ for the shower was found to be 2.10 $\pm$0.10 (Fig. \ref{fig:mag_dist}). The mean value is slightly above the upper limit of the range 1.87$<r<$2.01 given by \citet{Brown1996} during the years 1988-1994 but within the range 1.86$<r<$2.12 found by \citet{Jenniskens2016} for the years 1991-1994. These observations cover both the pre- and post activity of the meteor shower that coincided with the perihelion passage of the parent comet in 1992. The index was computed from 61 meteors brighter than +0 m. The light curve of a bright, double-flaring Perseid was computed following the method described in Section \ref{meteor_photometry} (Fig. \ref{fig:lightcurve}). The time step size $dt$ was 0.006 and 0.01 seconds for the un-shuttered and shuttered camera respectively. The gaps between the points represent the time intervals with no information owing to the rotating shutter.

Heliocentric orbits were computed for all 177 meteors in our sample. The median orbital elements of the 71 Perseid meteors were compared with the orbits found in five studies \citep{Kresak1970,Jopek2003,SonotaCo2009,Jenniskens2016} and the orbit of comet 109P/Swift-Tutle, parent comet of the Perseids \citep{Jenniskens2006Book} (Table \ref{table:table2}). In general, we find a good agreement in the radiant, speed, and orbital elements. One exception is in the semi-major axis, which is known to be very sensitive to variations in the velocity of a meteoroid \citep{Williams1996,Jenniskens1998}.

\section{Conclusions}
\label{conclusions}
We have presented SPOSH-Red, a software package for the data reduction of double-station meteor image data acquired by the SPOSH camera. The software extracts information about trajectories, heliocentric orbits and brightness levels of recorded meteors. 

The software was tested for simulated and real meteor data. We simulated different geometric configurations between a meteor shower and two observing sites. We suggest that such simulations can be used to assess the quality of the derived meteor trajectories for different camera network configurations and predicted meteor shower or outbursts. We expect that the results will greatly contribute to the planning of observing campaigns by finding the best location and orientation between the camera stations and predicted radiant position.

The software presented in this paper was developed to reduce data acquired by the SPOSH camera system. In the future we plan to provide a more generic version of the software package that can handle image datasets recorded by different camera systems.

The real meteor data used in this work is part of a large dataset comprising eight years of Perseids observations using SPOSH. The accumulated observing period spans more than 20 days within the activity period of the Perseids. During this period $>$ 15,000 single meteors have been recorded from both stations. The reduction software opens the opportunity to analyse the available unique SPOSH meteor data. A full analysis of the data focussing on the Perseid meteor shower will be presented in a forthcoming paper.

\section*{Acknowledgment}
The authors would like to thank the reviewer and editor for providing us with useful comments and detailed suggestions for the manuscript.

\bibliographystyle{aa} 
\bibliography{Per_v13}

\end{document}